\documentclass[
twocolumn,
twocolappendix,
tighten,
]{aastex63}
\hypersetup{
    linkcolor=[rgb]{0.6,0.3,0.2},
    citecolor=[rgb]{0.1,0.38,0.95},
    filecolor=cyan,
}
\usepackage{amsmath}
\usepackage{amstext}
\usepackage[T1]{fontenc}
\usepackage{apjfonts}

\usepackage{xspace}
\usepackage[encapsulated]{CJK}
\usepackage{ucs}
\usepackage{mleftright}
\usepackage{float}
\usepackage{booktabs}
\usepackage[all]{hypcap}

\newcommand{\cntext}[1]{\begin{CJK}{UTF8}{gbsn}#1\end{CJK}}

\graphicspath{{./}{figures/}}

\submitjournal{ApJ}
\shorttitle{CLASS Cloud observation}
\shortauthors{Li et al.}

\begin{document}
\title{CLASS Observations of Atmospheric Cloud Polarization at Millimeter Wavelengths}
\newcommand{\jhu}{The William H. Miller III Department of Physics and Astronomy, Johns Hopkins University, 3701 San Martin Drive, Baltimore, MD 21218, USA}
\newcommand{\ucsc}{Departamento de Ingenier\'{i}a El\'{e}ctrica, Universidad Cat\'{o}lica de la Sant\'{i}sima Concepci\'{o}n, Alonso de Ribera 2850, Concepci\'{o}n, Chile}
\newcommand{\villanova}{Department of Physics, Villanova University, 800 Lancaster Avenue, Villanova, PA 19085, USA}
\newcommand{\goddard}{NASA Goddard Space Flight Center, 8800 Greenbelt Road, Greenbelt, MD 20771, USA}
\newcommand{\uchicago}{Department of Astronomy and Astrophysics, University of Chicago, 5640 South Ellis Avenue, Chicago, IL 60637, USA}
\newcommand{\puci}{Instituto de Astrof\'isica, Facultad de F\'isica, Pontificia Universidad Cat\'olica de Chile, Avenida Vicu\~na Mackenna 4860, 7820436, Chile}
\newcommand{\pucc}{Centro de Astro-Ingenier\'ia, Facultad de F\'isica, Pontificia Universidad Cat\'olica de Chile, Avenida Vicu\~na Mackenna 4860, 7820436, Chile}
\newcommand{\argonne}{High Energy Physics Division, Argonne National Laboratory, 9700 S. Cass Avenue, Lemont, IL 60439, USA}
\newcommand{\upenn}{Department of Physics and Astronomy, University of Pennsylvania, 209 South 33rd Street, Philadelphia, PA 19104, USA}
\newcommand{\ucboulder}{Department of Astrophysical and Planetary Sciences, University of Colorado, 2000 Colorado Avenue, Boulder, CO 80309, USA}
\newcommand{\cfa}{Center for Astrophysics, Harvard \& Smithsonian, 60 Garden Street, Cambridge, MA 02138, USA}
\newcommand{\oslo}{Institute of Theoretical Astrophysics, University of Oslo, P.O. Box 1029 Blindern, N-0315 Oslo, Norway}
\newcommand{\MIT}{MIT Kavli Institute, Massachusetts Institute of Technology, 77 Massachusetts Avenue, Cambridge, MA 02139, USA}
\newcommand{\cepia}{CePIA, Astronomy Department, Universidad de Concepción, Casilla 160-C, Concepción, Chile}
\newcommand{\uiuc}{Department of Physics, University of Illinois at Urbana-Champaign, Urbana, IL 61801, USA}

\author[0000-0002-4820-1122]{Yunyang Li (\cntext{李云炀}\!\!)}
\affiliation{\jhu}
\correspondingauthor{Yunyang Li}
\email{yunyangl@jhu.edu}

\author[0000-0002-8412-630X]{John~W. Appel}\affiliation{\jhu}
\author[0000-0001-8839-7206]{Charles L. Bennett}\affiliation{\jhu}
\author[0000-0001-8468-9391]{Ricardo Bustos}\affiliation{\ucsc}
\author[0000-0003-0016-0533]{David T. Chuss}\affiliation{\villanova}
\author[0000-0002-7271-0525]{Joseph~Cleary}\affiliation{\jhu}
\author[0000-0002-0552-3754]{Jullianna Denes~Couto}\affiliation{\jhu}
\author[0000-0002-1708-5464]{Sumit Dahal}\affiliation{\goddard}\affiliation{\jhu}
\author[0000-0003-3853-8757]{Rahul Datta}\affiliation{\uchicago}\affiliation{\jhu}
\author{Rolando D\"unner}\affiliation{\puci}\affiliation{\pucc}
\author[0000-0001-6976-180X]{Joseph R. Eimer}\affiliation{\jhu}
\author[0000-0002-4782-3851]{Thomas~Essinger-Hileman}\affiliation{\goddard}
\author[0000-0003-1248-9563]{Kathleen Harrington}\affiliation{\argonne}\affiliation{\uchicago}
\author[0000-0001-7466-0317]{Jeffrey Iuliano}\affiliation{\upenn}\affiliation{\jhu}
\author[0000-0003-4496-6520]{Tobias~A. Marriage}\affiliation{\jhu}
\author[0000-0002-4436-4215]{Matthew~A.~Petroff}\affiliation{\cfa}
\author[0000-0001-5704-271X]{Rodrigo A. Reeves}\affiliation{\cepia}
\author[0000-0003-4189-0700]{Karwan Rostem}\affiliation{\goddard}
\author[0000-0001-7458-6946]{Rui Shi (\cntext{时瑞}\!\!)}\affiliation{\jhu}
\author[0000-0003-3487-2811]{Deniz A. N. Valle}\affiliation{\jhu}
\author[0000-0002-5437-6121]{Duncan J. Watts}\affiliation{\oslo}
\author[0009-0005-0983-986X]{Oliver F. Wolff}\affiliation{\uiuc}
\author[0000-0002-7567-4451]{Edward J. Wollack}\affiliation{\goddard}
\author[0000-0001-5112-2567]{Zhilei Xu (\cntext{徐智磊}\!\!)}\affiliation{\MIT}
\collaboration{24}{CLASS Collaboration}

\begin{abstract}
The dynamic atmosphere imposes challenges to ground-based cosmic microwave background observation, especially for measurements on large angular scales.
The hydrometeors in the atmosphere, mostly in the form of clouds, scatter the ambient thermal radiation and are known to be the main linearly polarized source in the atmosphere. 
This scattering-induced polarization is significantly enhanced for ice clouds due to the alignment of ice crystals under gravity, which are also the most common clouds seen at the millimeter-astronomy sites at high altitudes.
This work presents a multifrequency study of cloud polarization observed by the Cosmology Large Angular Scale Surveyor (CLASS) experiment on Cerro Toco in the Atacama Desert of northern Chile, from 2016 to 2022, at the frequency bands centered around 40, 90, 150, and 220~GHz.
Using a machine-learning-assisted cloud classifier, we made connections between the transient polarized emission found in all four frequencies with the clouds imaged by monitoring cameras at the observing site. 
The polarization angles of the cloud events are found to be mostly $90^\circ$ from the local meridian, which is consistent with the presence of horizontally aligned ice crystals. 
The $90$ and $150~\mathrm{GHz}$ polarization data are consistent with a power law with a spectral index of $3.90\pm0.06$, while an excess/deficit of polarization amplitude is found at $40/220~\mathrm{GHz}$ compared with a Rayleigh scattering spectrum. These results are consistent with Rayleigh-scattering-dominated cloud polarization, with possible effects from supercooled water absorption and/or Mie scattering from a population of large cloud particles that contribute to the $220~\mathrm{GHz}$ polarization.

\end{abstract}

\keywords{
    \href{https://astrothesaurus.org/uat/258}{Earth's clouds (258)};
    \href{https://astrothesaurus.org/uat/1278}{Polarimetry (1278)};
    \href{https://astrothesaurus.org/uat/322}{Cosmic microwave background radiation (322)}
}

\section{Introduction} \label{sec:intro}
Recent advances in cosmology from cosmic microwave background (CMB) observations greatly benefit from the use of ground-based telescopes \citep{aiola20,SPT3G-2021,BK-XV20} that complement space missions with high sensitivity polarization and small-scale intensity measurements.
CMB observations from the ground are limited by the atmosphere in many ways. 
Strong emission lines, mostly from oxygen and water vapor, define the transmission windows.
The turbulent and bulk motion of the atmosphere causes fluctuations in its brightness temperature on large time/angular scales, especially for inhomogeneously-mixed species like water vapor, which leads to long-term instabilities of CMB measurements \citep{Errard2015, Morris2021}.
Polarization measurements taking advantage of rapid modulation \citep{Johnson07-HWP, chus12vpm} are less sensitive to the unpolarized component of atmospheric turbulence, but the fluctuations could still degrade polarization sensitivity through instrument polarization.

Despite being mostly unpolarized \citep{battistelli2012}, the atmosphere is known to be circularly polarized due to the Zeeman effect from the magnetic dipole transition of the oxygen molecules in the Earth's magnetic field \citep{Lenoir1967, Lenoir1968,petroff20}.
This is a systematic issue for experiments with polarization modulation \citep{chus12vpm,nagy17} and presents a challenge to ground-based pursuits of astrophysical/cosmological circular polarization signatures \citep{King&Lubin2016,padilla20,eimer23}.

\begin{figure}
\centering
\includegraphics[width=1.0\linewidth, trim={0, 1.\linewidth, 0, 0.7\linewidth},clip]{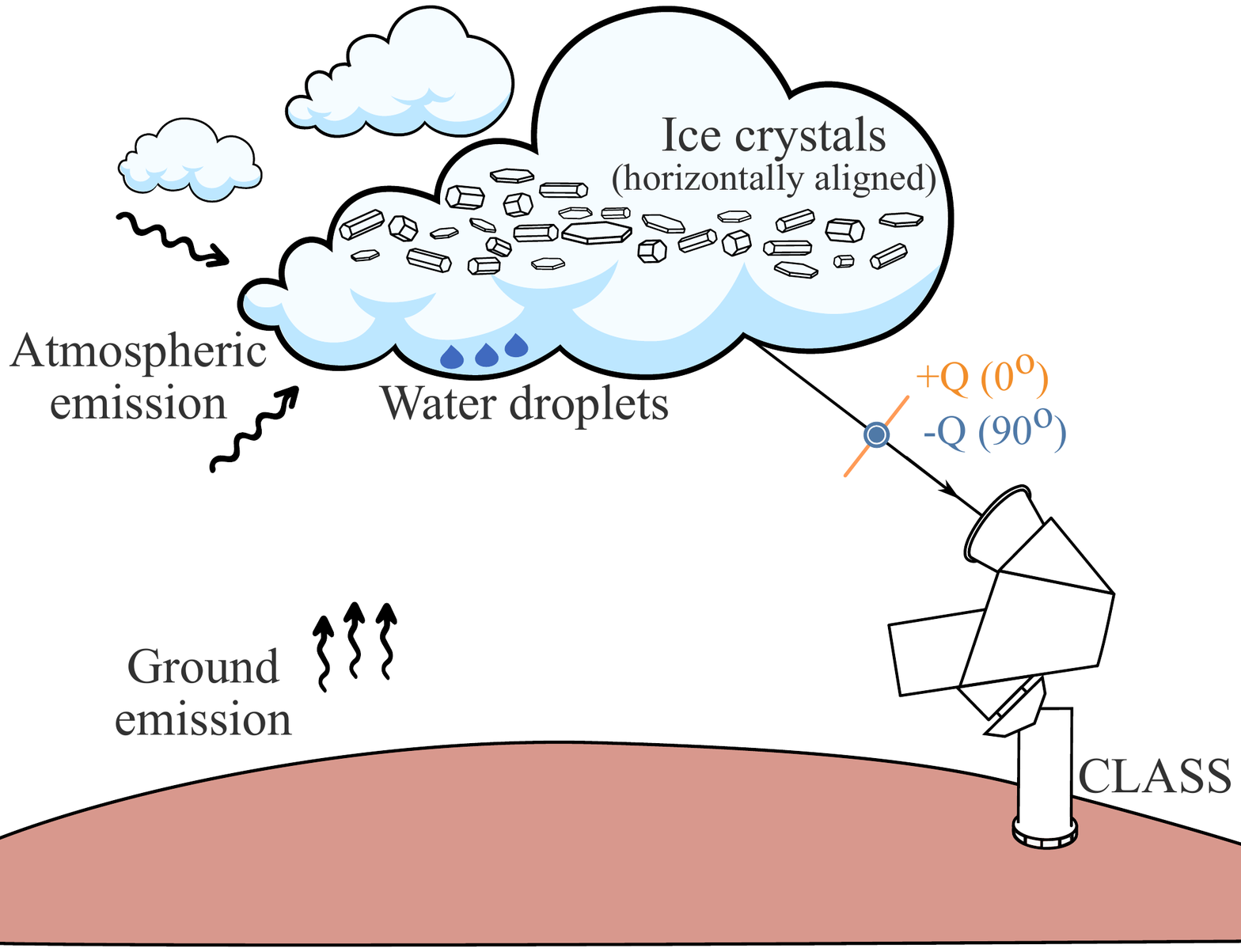}
\caption{Illustration of cloud polarization. Clouds containing water droplets and horizontally aligned ice crystals scatter thermal radiation from the ground and the atmosphere, directing linearly polarized radiation into the telescope. 
The two orthogonal linear polarization states are labeled; they correspond to the $Q$ ($0^\circ$, in orange) and $-Q$ ($90^\circ$, i.e., horizontal, in blue) Stokes parameter evaluated in the horizontal frame of the telescope.
Clouds are depicted as cumulus clouds instead of cirrus for their aesthetic appeal.
Objects are not to scale. 
}
\label{fig:cartoon}
\end{figure}

At millimeter wavelengths, linear polarization can be generated from unpolarized radiation via scattering by a hydrometeor in its liquid or ice phase. 
Polarization from the cloudy atmosphere was first detected at $100~\mathrm{GHz}$ by \cite{Troitsky&Osharin2000} and modeled as the scattering of the upwelling ground emission and atmospheric emission by the cloud particles.
Similar observations have been made across millimeter wavelengths from both the ground \citep{Troitsky+2003,Troitsky+2005} and remote-sensing satellites \citep{chepfer1999, Prigent+2005, Davis2005, Defer+2014, Gong&Wu2017}. 
The cloud polarization from ground-based observations is mostly horizontally aligned (i.e., negative Stokes $Q$ parameter. See Figure \ref{fig:cartoon} for a schematic illustration) albeit with some exceptions of $+Q$, which are attributed to spherical cloud particles \citep{Troitsky+2003} or mixed-phase cloud \citep[e.g., melting ice crystals,][]{Troitsky&Osharin2000, Troitsky+2005}.
For non-spherical cloud particles (e.g., ice crystals), several mechanisms are at play to align the particle orientations and thus enhance polarization. 
\cite{Vonnegut1965} points out that ice crystals could be vertically aligned by a discharge event in the cloud \citep{Hendry&McCormick1976,Prigent+2005}.
More commonly, ice crystals are found to be horizontally aligned \citep{magono1953, ono1969} since a hydrodynamic force acts on ice particles as they settle, tending to orient them with their broadside to the fall direction to maintain maximum air resistance \citep{klett1995orientation}.
Due to the counter-effect from turbulence \citep{breon2004,Hashino2014, zeng2023}, the concentration of horizontal alignment is often low, and
the alignment is most frequently found in warm clouds ($\gtrsim-20^\circ\mathrm{C}$), with a typical occurrence of about 20--80\% depending on the temperature and the ice crystal habits \citep{westbrook2010doppler,Noel&Chepfer2010,Stillwell2019}.

The impact on CMB observations from polarized cloud emission is found by \cite{Pietranera2007} to be twofold, namely, the backward scattering of the ground thermal emission and the forward scattering of the sky signal. 
While the latter distorts the CMB polarization signal, the former is a more substantial issue due to the large difference between the ground temperature (which sources the polarized cloud reflection) and the CMB polarization amplitude. 
The POLARBEAR team \citep{Takakura2019} made the first connection between the cloud events seen in camera images during the daytime and burst signals in their $150~\mathrm{GHz}$ detectors. The polarization of the bursts is consistent with the scattering of cirrus clouds composed of horizontally aligned ice crystals.

The Cosmology Large Angular Scale Surveyor \citep[CLASS,][]{essinger-hileman14spie} is a CMB polarization experiment located on Cerro Toco in the Atacama Desert that aims to measure polarization on large angular scales from the ground.
Here we provide an assessment of the cloud signal in multifrequency CLASS observations at 40, 90, 150, and 220~GHz between 2016 August and 2022 May. Table~\ref{tab:bandpass-info} summarizes the effective frequencies of the sources \citep[assuming a Rayleigh scattering spectrum $T_\mathrm{RJ}\propto \nu^4$,][]{dahal22} and the start date of each detector array.
The polarization modulation technique employed by CLASS enables both linear and circular polarization sensitivity. 
The fast surveying of the sky (scan through $36\%$ of the half-sky twice every 10~minutes), accompanied by all-day optical image monitoring, offers a holistic view of the cloud conditions on Cerro Toco.

This paper begins with a review of the radiative properties of clouds in Section~\ref{sec:physics}, and describes optical cloud identification from camera images in Section~\ref{sec:optical-detection} and microwave polarization observations in Section~\ref{sec:radio-detection}. 
Section~\ref{sec:population} summarizes the population analyses on cloud polarization, and we conclude in Section~\ref{sec:conclusion}.

\begin{deluxetable}{ccccc}
    \tablecaption{CLASS multifrequency information \label{tab:bandpass-info}}
    \tablehead{\colhead{Band} & \colhead{40} & \colhead{90} &\colhead{150}& \colhead{220}
    }
    \startdata
    \specialrule{.05em}{0em}{0em}
    $\nu_\mathrm{eff}\,\mathrm{(GHz)}$ & 38.5 & 94.2 & 148.2 & 220.0 \\
    Start Date & 2016-08 & 2018-06 & 2019-10 & 2019-10 \\
    \specialrule{.05em}{0em}{0em}
    \enddata
    \tablecomments{The numbers are calculated assuming the fiducial Rayleigh scattering spectrum.}
\end{deluxetable}

\section{Cloud physics} \label{sec:physics}
\subsection{Cloud Characterization}

\begin{figure}
\centering
\includegraphics[width=1.0\linewidth, trim={0.04\linewidth, 0.04\linewidth, 0.02\linewidth,  0},clip]{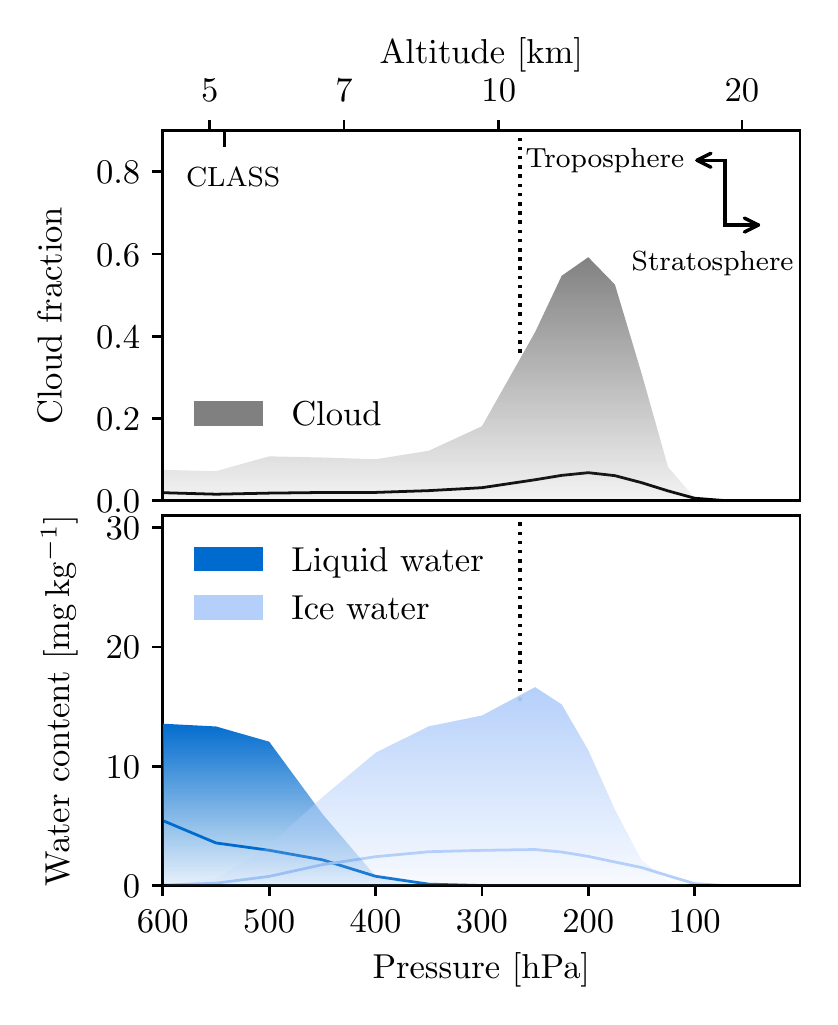}
\caption{Cloud fraction and water content profile near the CLASS site from 2016 August to 2022 May derived from ERA5 \citep{ERA5} per-pressure level hourly data. 
The pressure levels are converted into altitudes above sea level (top axis) using the annually-averaged atmosphere profile from \cite{merra2} and the hydrostatic equation.
\textit{Top:} cloud cover fraction as a function of atmosphere pressure (altitude). 
This quantity is the proportion of a grid box covered by cloud in the ERA5 model.
\textit{Bottom:} the specific cloud water content in liquid (darker blue) and ice (lighter blue) phase. 
The area enclosed under the curves (in linear pressure axis) is proportional to the Liquid- (Ice-) water path.
In both panels, the solid curves show the mean value at given pressure levels; the shaded regions fill between the $50$ and $95$ percentile of the distribution. 
Due to aridness at the queried region, the $50$ percentiles are essentially zero.
The vertical dotted line marks the altitude where the median air temperature is $-40~\mathrm{^\circ C}$, above which (altitude) homogeneous nucleation occurs. 
}
\label{fig:cloud_profile}
\end{figure}

Altitude is the main factor in determining the cloud composition and its classification.
At 5200~m on Cerro Toco, the predominant types are \emph{high-level clouds} (above 6~km in the tropical region, e.g., \textit{Cirrus, Cirrostratus, Cirrocumulus}, sub-classified based on the morphology) that are almost entirely composed of ice crystals and \emph{mid-level clouds} (up to 8~km, e.g., \textit{Altostratus, Altocumulus}) that may contain both ice and (supercooled) water droplets. 

The water content in the atmosphere is often quantified by the precipitable water vapor (PWV) and liquid/ice water path (L/IWP), which measure the total mass of the water per unit area in liquid/solid phases.
As one of the exceptionally arid sites for mm-astronomy, the Chajnantor area experiences a median $1~\mathrm{mm}$ PWV \citep{Cortes2020}. 
Water in the hydrometeor form is long-tail distributed, with LWP and IWP below $10^{-3}~\mathrm{kg\,m^{-2}}$ 80\% of the time but could rise above $10^{-2} ~\mathrm{kg\,m^{-2}}$ for $10\%$ of the time \citep[][austral summer months excluded]{Kuo2017}.
This is also consistent with the typical $10^{-2}~\mathrm{kg\,m^{-2}}$ IWP found for cirrus clouds \citep{davis2007comparisons, Austin+2009}.

Figure \ref{fig:cloud_profile} shows the vertical profile of the cloud fraction and liquid and ice water content within the atmosphere above the CLASS site. 
The data near the CLASS site, within a $0.25^\circ\times0.25^\circ$ pixel centered at $67.8^\circ\,\mathrm{W}, 23.0^\circ\,\mathrm{S}$, were queried from the global reanalyses ERA5 \citep{ERA5}.
The annual averaged cloud profile shows a bimodal distribution; mid-level clouds with liquid-water constituents are $\lesssim 2~\mathrm{km}$ above the ground ($\lesssim7~\mathrm{km}$ above sea level), and ice clouds are higher in altitude, at the boundary of the troposphere.
We note that the water content and cloud fraction distributions are highly skewed and that the records are zero most of the time.

\subsection{Rayleigh Scattering}\label{sec:Rayleigh-scattering}
In the Rayleigh limit where the scatterer size is small compared to the wavelength, the electric field component of the scattered radiation $\mathbf{E}_o$ at direction $\mathbf{r}$ from the scatterer follows\footnote{In most of the cases (Figure \ref{fig:cloud_profile}), the cloud distance is well within the far-field region of the telescopes ($\gg 200~\mathrm{m}$ for the $220~\mathrm{GHz}$ telescope).}
\begin{equation}
\mathbf{E}_o = -\frac{V}{r^3}\left(\frac{2\pi\nu}{c}\right)^2\mathbf{r}\times(\mathbf{r}\times\boldsymbol\alpha\mathbf{E}_i),
\end{equation}
where $V$ is the scatterer volume; $\nu$ is the frequency of the incoming radiation $\mathbf{E}_i$; $c$ is the speed of light. 
The polarizability tensor, $\boldsymbol{\alpha}$, depends on the dielectric properties and shape of the particle. Water droplets are often spherical, whereas ice crystals have a variety of habits ranging from regular shapes including quasi-spheroids, elongated columns, thin plates, and branched dendrites (Figure \ref{fig:cartoon}) depending on the temperature and vapor supersaturation \citep{libbrecht2017physical} to highly asymmetric aggregates \citep{lawson2019}. 
For simplicity, we model all particle types as spheroidal;
following notations in \cite{Takakura2019}, the polarizability tensor is parameterized as
\begin{equation}
\boldsymbol{\alpha}=\frac{\epsilon-1}{4\pi}\left[\mathbf{I}+(\epsilon-1)\boldsymbol\Delta\right]^{-1}\label{eq:permittivity},
\end{equation}
where $\epsilon$ is the relative permittivity; $\mathbf{\Delta} =\mathrm{diag}\{\frac{1-\Delta_z}{2}, \frac{1-\Delta_z}{2}, \Delta_z\}$ is the depolarization matrix for spheroidal scatterers (with $\Delta_z$ less or greater than $1/3$ for prolate or oblate crystals, respectively).

\begin{figure}
\centering
\includegraphics[width=1.0\linewidth, trim={0.0\linewidth, 0, 0.0\linewidth,  0},clip]{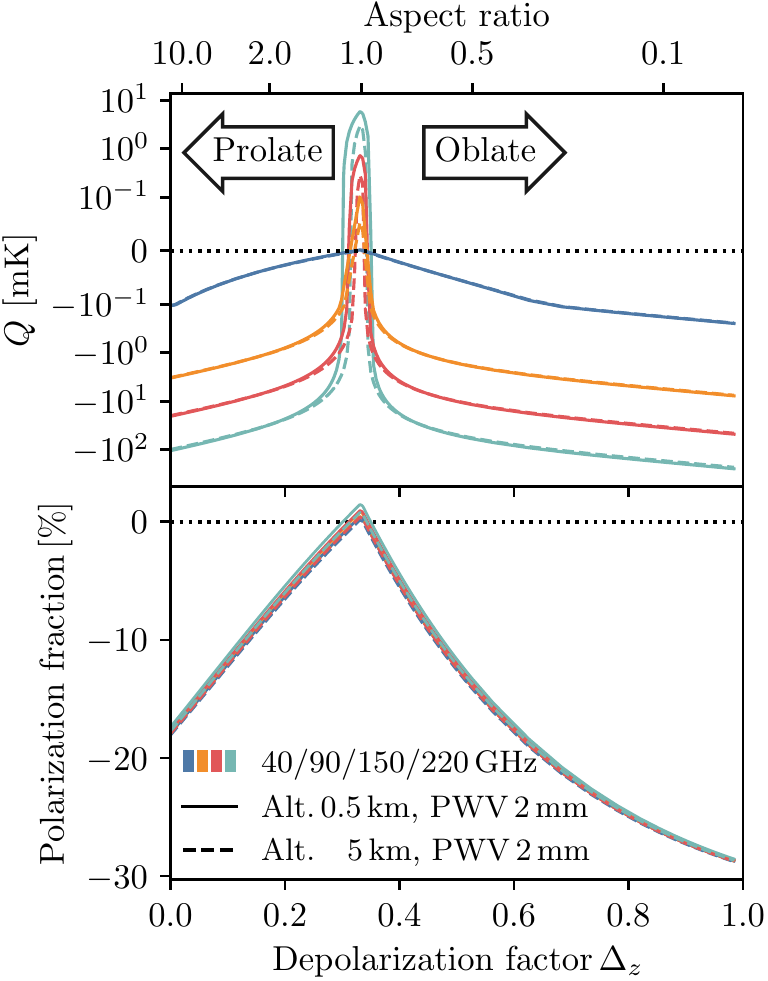}
\caption{Polarized signal from Rayleigh scattering of thermal emission from the ground and the atmosphere for different cloud particle shapes. Assuming homogeneous spheroidal cloud particles with uniform dielectric properties, the particle shape is parameterized by its aspect ratio, or equivalently, the depolarization factor $\Delta_z$. The linear polarization from an ensemble of randomly oriented cloud particles with the short axis aligned vertically is computed for each of the four CLASS frequency bands (color-coded), for different cloud heights above the ground (5200$\,$m above sea level), with a total $2~\mathrm{mm}$ PWV. 
The cloud height and PWV determine the optical loading above the cloud (Equation \ref{eq:brightness}), and impact the polarization from the forward scattering of atmospheric emission, especially at higher frequencies.
\textit{Top:} the Stokes $Q$ polarization signal in units of brightness temperature for a reference optical depth $\tau_{\nu}$ equivalent to $\mathrm{IWP}=10~\mathrm{g\,m^{-2}}$ and particle diameter $D=100~\mathrm{\mu m}$ (geometric mean of the three axes). \textit{Bottom:} the corresponding polarization fraction. Signed values are used to distinguish the sign of Stokes $Q$ polarizations.}
\label{fig:cloud_model}
\end{figure}

For cloud reflection, we define the coordinates centered on the scatterer and evaluate the radiation in terms of Stokes parameters. 
The incoming radiation is assumed to be unpolarized: $\mathbf{S}_{i}=[T(\theta), 0, 0, 0]^\mathrm{T}$, with intensity profile
\begin{equation}\label{eq:brightness}
    T(\theta)=
    \begin{cases}
    T_\mathrm{g},&  \theta< \theta_\mathrm{h}\\
    T_a(\theta, \nu),              & \theta\geq \theta_\mathrm{h}
\end{cases},
\end{equation}
where $\theta$ is the zenith angle;  $\theta_\mathrm{h}\approx\sqrt{2h/R_\mathrm{e}}$ is the astronomical horizon for clouds at altitude $h$ above the ground ($R_\mathrm{e}$ is the radius of the Earth); $T_\mathrm{g}$ is the ground temperature, and $T_a(\theta, \nu)$ is the brightness temperature profile of the atmosphere. 
This setup is illustrated in Figure \ref{fig:cartoon}.
For spherical particles with $\boldsymbol{\alpha}\propto(\epsilon-1)/(\epsilon+2)\mathbf{I}$, the Stokes parameters from cloud reflection evaluated in the telescope frame can be expressed as
\begin{equation}
\mathbf{S}_{o}(\nu) = \tau_\nu\int\mathbf{R}(\theta,\phi, \delta)\cdot\mathbf{M}(\theta,\phi, \delta) \mathbf{S}_{i}(\theta, \nu)\mathrm{d}\,\mathbf{\cos\theta}\mathrm{d}\,\mathbf{\phi}\label{eq:scattering-formula}
\end{equation}
where $\mathbf{S}_{o}(\nu)$ is the radiation received by the telescope at $(-\delta, 0)$, with $\delta$ being the telescope pointing elevation; Matrix $\mathbf{R}$ rotates Stokes vectors into the telescope frame from the scattering frame in which the phase matrix $\mathbf{M}$ is evaluated. 
The column of $\mathbf{M}$ for unpolarized incoming radiation is $(1/2)[(\cos^2\Theta+1), (\cos^2\Theta-1), 0, 0]^\mathrm{T}$, where $\Theta(\theta, \phi, \delta)$ is the scattering angle.
The Rayleigh scattering optical depth is
\begin{align}
\tau_\nu &= \int \sigma_\mathrm{sca} n(\mathbf{r})\mathrm{d}\mathbf{r} \notag\\
 &= \int4\pi \left(\frac{2\pi\nu}{c}\right)^4\left(\frac{D}{2}\right)^6\left|\frac{\epsilon-1}{\epsilon+2}\right|^2 n(\mathbf{r})\mathrm{d}\mathbf{r} \label{eq:tau-formula}\\
&\approx1.2\!\times\!10^{-5}\left(\frac{\nu}{100~\mathrm{GHz}}\right)^4\!\left(\frac{D}{100~\mathrm{\mu m}}\right)^3\!\left(\frac{\mathrm{IWP}}{10~\mathrm{g\,m^{-2}}}\right)\label{eq:tau-value},
\end{align}
where $\sigma_\mathrm{sca}$ is the Rayleigh scattering cross section; $n(\mathbf{r})$ is the number density of cloud particles with diameter $D$.
The typical crystal size in cirrus is $100~\mathrm{\mu m}$, although the distribution is widespread \citep{lawson2019, Austin+2009}. 
At mm-wavelengths the dielectric permittivity for ice is, $\epsilon\simeq3.2$, \citep{Warren&Brandt2008}, and being primarily of interest in this setting (see Figure \ref{fig:cloud_model}) is adopted in evaluating Equation \ref{eq:tau-value}. For liquid water, the permittivity has a larger magnitude and can be approximated by a Debye dielectric function \citep[][also see discussion in Section \ref{ssec:spectra}]{Ellison2007, turner2016improved}.

Figure \ref{fig:cloud_model} shows the total linear polarization (Stokes $Q$, since $U$ polarization in the horizontal frame is zero under parity symmetry) and the polarization fraction, observed at $\delta=45^\circ$ elevation, from horizontally aligned spheroidal cloud particles with random azimuthal orientations as a function of the depolarization factor $\Delta_z$ (also converted to the aspect ratio of the spheroids assuming uniform permittivity). 
The absolute polarized emission is evaluated assuming an optical depth corresponding to a $10~\mathrm{g\,m^{-2}}$ LWP and a particle size $100~\mathrm{\mu m}$ (geometric mean diameter). The atmospheric brightness profile $T_a(\theta, \nu)$ is computed with \texttt{am} \citep{am}, using the annually-averaged atmosphere profile at the Chajnantor site compiled from the reanalysis of MERRA-2 \citep{merra2} and the CLASS bandpass \citep{dahal22}. 
The linear polarization fraction is highest for non-spherical scatterers with extreme aspect ratios and peaks at $\sim30\%$ for oblate crystals at $45^\circ$ elevation. 
For spherical particles, the $Q$ polarization could be positive due to the thermal emission from the atmosphere, which is more prominent at higher frequencies, and for clouds at lower altitudes where the air temperature and water vapor above the cloud are higher. 
This is the main modeling difference from \cite{Takakura2019}, which did not consider the brightness temperature contribution from the atmosphere.
Ignoring the forward scattering of atmospheric emission in Equation \ref{eq:brightness} and solving for the linear polarization component with Equation \ref{eq:scattering-formula}, we find negative $Q$ polarization for all particle shapes, with the lowest amplitude from spherical scatterers:
\begin{equation}\label{eq:simple-solution}
Q =-\frac{\pi\tau T_\mathrm{g}}{2}\cos^2{\delta}\sin\theta_\mathrm{h},
\end{equation}
and the polarization fraction $\lesssim 1\%$ for $\delta=45^\circ$. This agrees with the derivation in \cite{Takakura2019}.

The derivation here only considers single scattering and a full radiative transfer calculation is needed to self-consistently account for the Rayleigh scattering and contributions from absorption/emission of the atmosphere; nevertheless, the result is consistent with the full radiative transfer model \citep{Troitsky&Osharin2000}.

\subsection{Mie Scattering}
At wavelengths comparable to the size of the cloud particles, the Rayleigh scattering approximation breaks down and the full treatment using the Mie theory is needed.
The Mie solution differs from the Rayleigh approximation in three major aspects: (1) the resonance effect related to the size and refractive index of the scatterer is considered, which yields complex frequency dependency and is generally shallower than the Rayleigh scattering spectrum $\propto\nu^4$; (2) the forward scattering is enhanced compared to other directions; (3) the scattering phase matrix contains non-zero linear to circular polarization coupling terms.

\begin{figure}
    \centering
    \includegraphics{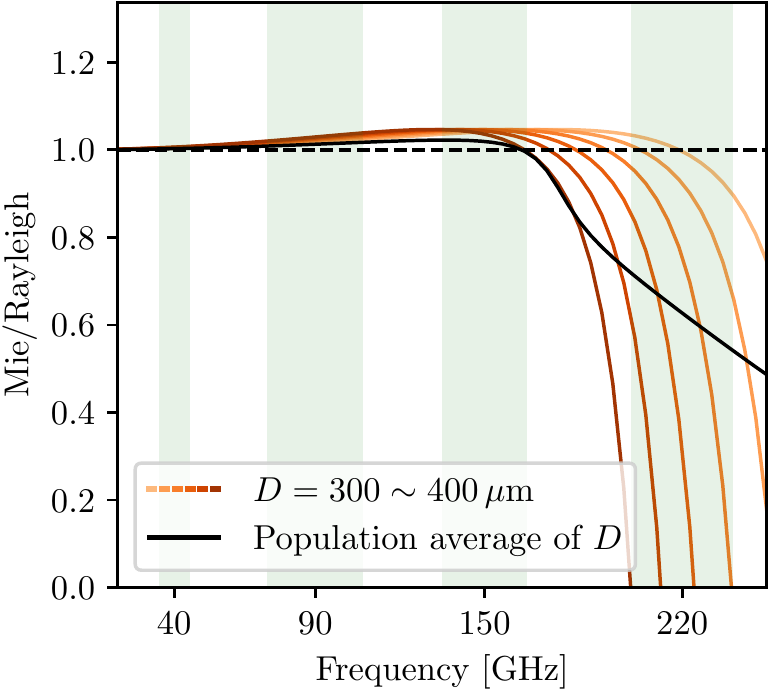}
    \caption{Cloud polarization spectrum ratio between the Mie solution and the Rayleigh scattering approximation for spherical ice particles. 
    The colored lines show the ratio evaluated for a homogeneous population with particle diameter ranging from $300$ to $400~\mathrm{\mu m}$. 
    The black curve shows the average result from a population of ice particles with log-normal size distribution at $-70\mathrm{^\circ C}$ \citep{Austin+2009}. 
    CLASS observation frequencies and bandwidths are shown as the shaded vertical bands.
    }
    \label{fig:mie-spectrum}
    \end{figure}

The first two points alter the cloud reflection signal. 
As a simplified model, we consider the cloud composition as homogeneous spherical particles and ignore the atmospheric emission in Equation \ref{eq:brightness}. 
The Rayleigh solution for polarization is given in Equation \ref{eq:simple-solution}, and we compute the corresponding result for Mie scattering using \texttt{miepython}.\footnote{\url{https://github.com/scottprahl/miepython}}
The results are shown in Figure \ref{fig:mie-spectrum} as the ratio of the linear polarization spectrum from the Mie solution and the Rayleigh scattering approximation.
The polarization signal at around $220~\mathrm{GHz}$ is suppressed in the Mie solution in a way that depends sensitively on particle size in the $300$ -- $400~\mathrm{\mu m}$ range.
While the majority of ice crystals are below this threshold, the size distribution is long-tailed \citep{McFarquhar&Heymsfield1997, Heymsfield+2002,Austin+2009}, and the larger particles can contribute significantly to the polarization signal (roughly $\propto D^6$).
Assuming the size distribution of ice crystals with log-normal mean size $46~\mathrm{\mu m}$ \citep[corresponding to the best-fit parameters at $-70~\mathrm{^\circ C}$,][]{Austin+2009} and standard deviation $0.226$, the black curve shows the population-averaged spectrum ratio. 
While Rayleigh scattering is a good approximation of cloud polarization at frequencies below $150~\mathrm{GHz}$, the Mie scattering effect cannot be ignored for higher frequencies. 
For realistic high cloud composition, the cloud polarization signal at the $220~\mathrm{GHz}$ band is a factor of a few smaller compared to the Rayleigh solution and is comparable to the signal at $150~\mathrm{GHz}$.

In the Rayleigh scattering regime, circular polarization from directly scattering atmospheric circular emission \citep[][]{petroff20} is negligible due to parity. 
For Mie scattering, the coupling between linear and circular polarization \citep{Bohren&Huffman1983}
can convert the linear polarization induced by preceding scatterings into circular polarization \citep{kawata1978,slonaker05}. 
However, the required multiple scattering is improbable given the low scattering optical depth. 

\section{Cloud Measure} \label{sec:optical-detection}
\subsection{Camera Images}
\begin{figure}[ht!]
\begin{center}
\includegraphics{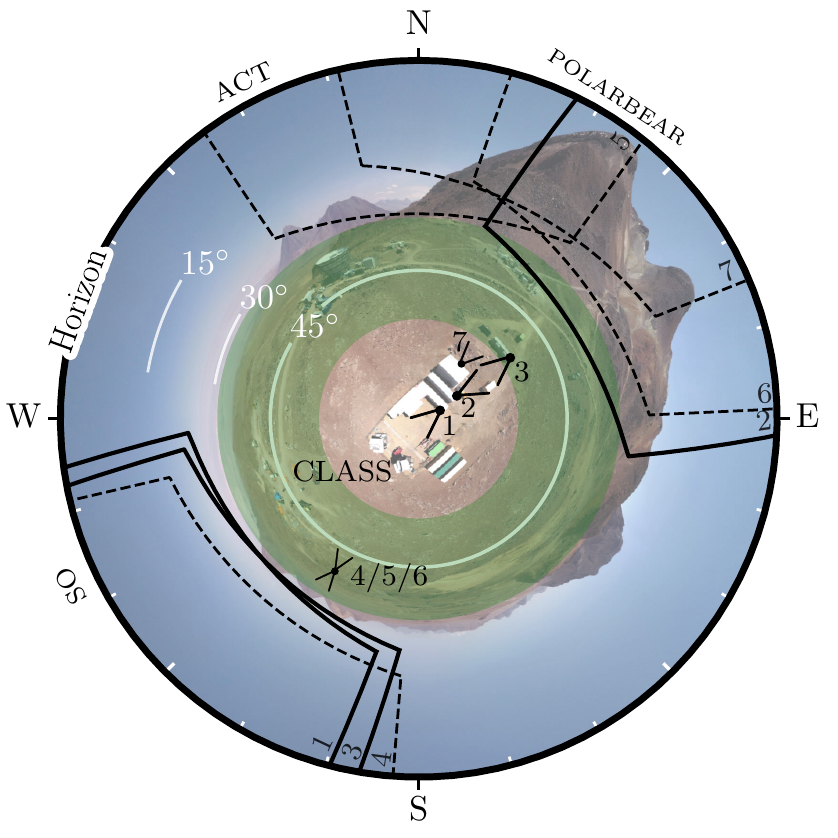}   
\end{center}

\caption{Aerial view of the CLASS site. 
The background image shows the CLASS site and its surrounding terrain; the two CLASS telescope mounts are slightly off-center. 
The direction toward neighboring CMB experiments on Cerro Toco: POLARBEAR, Atacama Cosmology Telescope (ACT), and the Simons Observatory (SO) are also labeled. 
Site camera locations and pointings are shown as black wedges, and their fields of view are delineated by the black border lines (solid borders for main cameras 1--3 and dashed ones for cameras that are daytime only). 
The shaded green region shows the footprint of the CLASS telescopes during the normal constant-elevation scan at $45^{\circ}$. 
Note that the telescope and camera FoVs are plotted in \textit{sky} coordinates with the \textit{zenith} at the center of the image, but the projection of the ground image centers at the \textit{nadir} direction; therefore, the region enclosed by the site camera border lines do not reflect the actual view of the cameras (the inner border lines are the top edge of the camera images)---only the azimuthal information matches the sky coordinates.\label{fig:camera-pano}
}
\end{figure}

CLASS has deployed several cameras at its observing site to monitor telescope operations. 
Some of these cameras have a large fraction of sky coverage, which can be used for cloud detection.
These images are taken at a 10-minute cadence, approximately the time for the telescope to complete a $720^\circ$ scan in azimuth. 
Figure~\ref{fig:camera-pano} shows the positioning and pointing directions of the seven cameras at the site and the landscape around the CLASS telescopes.
The three primary cameras (labeled ``1--3'') are functional during daytime (colored) and nighttime (grayscale), whereas the rest only have daytime data due to the limited shutter speed at night, and were installed in the middle of the survey.
For each stable period of each camera pointing, we built the world coordinate system (WCS) for the images. For cameras 1--3, this is accomplished by solving the star positions with \texttt{astrometry.net} \citep{astrometrynet}, whereas for daytime-only cameras, the camera models were solved from the terrain shapes, with geographical information from Google Earth. 
The fields of view (FoVs) of the cameras are shown as black wedges (solid and dashed) in Figure \ref{fig:camera-pano}, which have minimal intersection with the sky swept by the telescopes (green shades). 
Therefore, the connections we make in the following analyses have to assume similar cloud conditions seen by the telescopes and the cameras.

The cloud cover in each of the site images was estimated using a convolutional neural network (CNN) classifier. The details of the implementation and the validation of the classifier are described in Appendix~\ref{sec:cnn}.
The CNN classifier labels each of the $240 \times 240$ pixel sub-regions in an image as one of the three classes: \textit{cloud}, \textit{clear}, and \textit{obscured} (by the ground, the telescope, or celestial objects that prohibit a definitive cloud identification), and the cloud measure is defined as the ratio of sub-regions classified as cloud versus the total number that is not obscured (cloud + clear). 
Combined with the WCS of each site image, the cloud measure can be reported with finite angular resolution. In the rest of this work, we use the spatial average (over the sky covered by all cameras) of the cloud measure above $10^\circ$ elevation ($\mathrm{CM}_{10}$) as a proxy of the cloud condition at the site.

\subsection{Cloud Conditions at Cerro Toco}
\begin{figure*}
\centering
\includegraphics{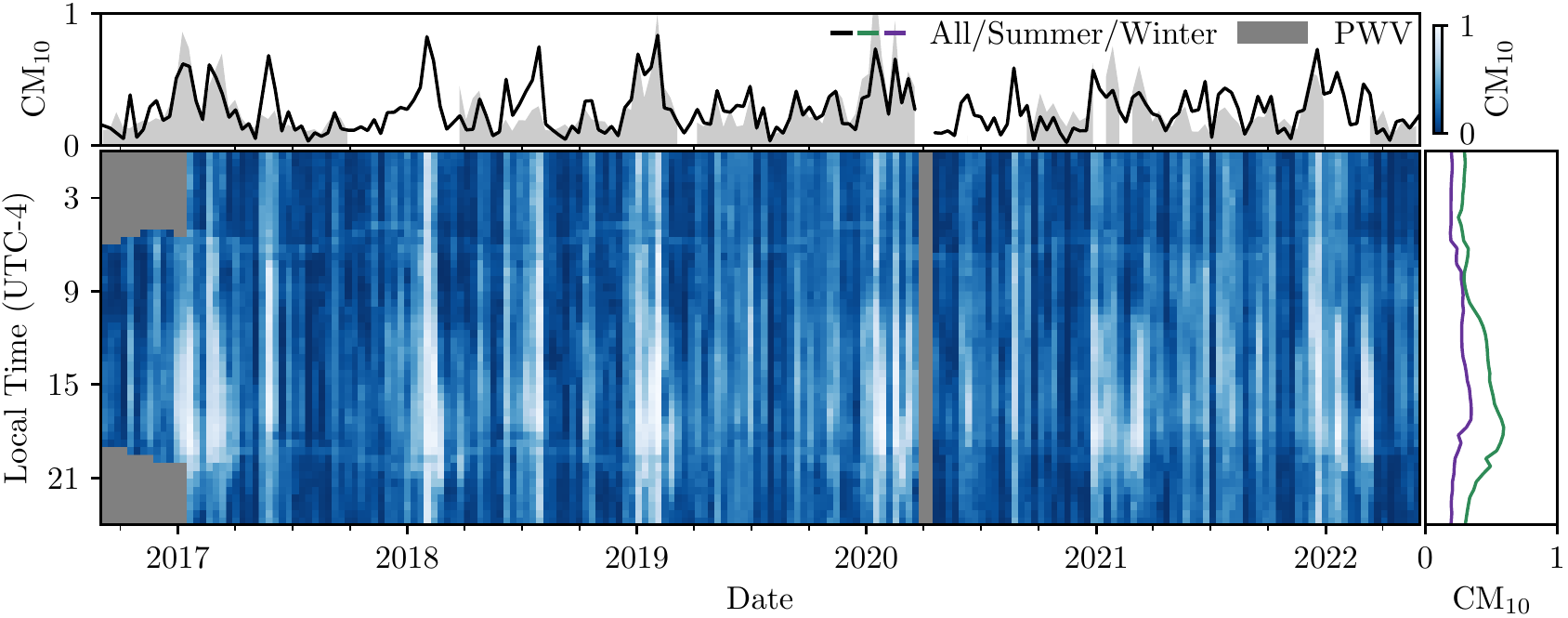}
\caption{Cloud cover fraction above elevation $10^\circ$ ($\mathrm{CM}_{10}$) measured by the CLASS site cameras from 2016 August to 2022 May. 
The nighttime data prior to 2017 January are discarded due to the low exposure time set for the cameras.
\textit{Main}: histogram of $\mathrm{CM}_{10}$ in local time and date. 
\textit{Top}: the annual variation. The PWV measurements during the same time period are shown as shaded curves, with arbitrary scaling. 
The gray shades break when the APEX weather data are not available.
\textit{Right}: the season-averaged diurnal variation during Austral summer (December, January, February, in green) and Austral winter (June, July, August, in purple).}
\label{fig:cloud-annual}
\end{figure*}

The high-cadence all-day monitoring during the six-year survey of CLASS provides an assessment of the cloud conditions at the site on Cerro Toco.
Figure \ref{fig:cloud-annual} shows the $\mathrm{CM}_{10}$ statistics from 2016 through 2022. 
Over the course of a day, the cloudiness begins increasing after midday and peaks around dusk; this agrees with the diurnal cycle of cirrus and cumulonimbus clouds over land \citep{Eastman2014,Feofilov2019}, which develop during the afternoon due to the mixing of the boundary layer and the increase in thermal convection. 
This variation is steeper at Summer times when the convection is most efficient and peaks around January, consistent with the record from POLARBEAR \citep{Takakura2019} at the same site.
The annual variation is also evident and tracks well with the PWV measurement from the APEX weather station.\footnote{\url{https://www.apex-telescope.org/ns/weather-data}}
The apparent yearly modulation of the cloudiness around twilight reflects the synchronous shift of the diurnal peak of cloud cover with solar motion; however, part of this is the systematic uncertainty due to the higher false-positive cloud detection caused by aerosol scattering around the Sun. 

The average cloud cover at Cerro Toco is around $32\%$ during daytime and $20\%$ during nighttime, modulated by a seasonal variation around 50\%, consistent with the cloud detection rate of daytime images taken from the neighboring POLARBEAR site \citep{Takakura2019}.

\section{Cloud Signal in Microwave}\label{sec:radio-detection}
\subsection{CLASS Instruments and Polarimetry}\label{sec:class-polarimetry}

CLASS observes at a constant $45^{\circ}$ elevation and scans continuously in azimuth by $720^\circ$ before turning around at a rate of $1$--$2^\circ\,\mathrm{s}^{-1}$. 
Polarization is measured through the variable-delay polarization modulator \citep[VPM,][]{chus12vpm,harrington18spie,harrington21}, which modulates between the linear polarization (Stokes $U$ in the telescope coordinates) and circular polarization at around $10~\mathrm{Hz}$.
For cosmological observations, the linear polarization angle coverage is achieved by the rotation of the Earth and the telescope boresight rotation from $-45^\circ$ to $45^\circ$ at $15^\circ$ increments every day. 
For a cloud transient event, however, the VPM is only sensitive to one polarization direction --- e.g., $\pm$ Stokes $Q$ for $\mp45^\circ$ boresight and Stokes $\pm U$ for $0^\circ$ boresight.
Nevertheless, polarimetry is still possible since the $\sim30^\circ$ azimuthal differential pointing of the detectors on the focal plane translates into a $\sim 15^\circ$ spread in position angle on the sky at $45^\circ$ elevation.
This finite position angle coverage permits the determination of the polarization angle from a single scan, albeit with lower precision compared to that from rapid modulation between the two linear polarization states through a continuously rotating half-wave plate \citep[e.g.,][]{Takakura2019}.

\subsection{Data Selection}
The radiometer data analyzed in this work are taken from the CLASS survey from 2016 August to 2022 May. 
As the 90~GHz and 150/220~GHz receivers were deployed for less time, their data covered a smaller date range.
The beginning date of each frequency is summarized in Table \ref{tab:bandpass-info}.
For the purpose of investigating the clouds, the data selection is less stringent compared to that for cosmological analysis.
The data were chunked into \textit{spans} defined by a segment of uninterrupted observations. 
Typically, a span is a day long and is only separated by the detector calibration \citep{appel22} and boresight rotation change at mid-day.
Similar to the data selection described in \cite{Li23}, data with instrument failure (VPM, cryostats, mount, readout system) or abnormal detector response (constant detector output or excessive flux jumps) were discarded. 
Data were masked when the telescope scans close to the Sun or the Moon. No other source avoidance was made.
Other environmental conditions such as the high PWV, and high cloud fraction were not used as selection criteria to avoid rejecting the cloud signals.

\subsection{Azimuthal maps}\label{sec:map-making}
To find cloud signals among other burst-like events in the data, and to obtain better polarization angle measurements, we made polarization maps combining multiple detectors from the focal plane for each of the Stokes parameters.

The linear ($u$) and circular ($v$) polarization signals were obtained by demodulating the time stream with the VPM transfer function \citep{harrington21,Li23}:
\begin{align}
    \
    \begin{bmatrix}
           u \\
           v
    \end{bmatrix} 
    &= \mathfrak{L}\left[\mathbf{M}^T\mathbf{M}\right]^{-1} \mathfrak{L}\left[ \mathbf{M}^T\mathbf{D}\right],
\label{eq:demod_solution}
\end{align}
where $\mathbf{D}$ is the raw time stream modulated by the VPM, $\mathbf{M}$ is the VPM modulation transfer function, and $\mathfrak{L}$ is the low-pass filter defined by the signal bandwidth given the beam size and scanning speed.
As the VPM modulates, its synchronous signal is also modulated at around $10~\mathrm{Hz}$ and so contributes to the single-detector demodulated linear polarization as a baseline offset\footnote{This synchronous signal drifts slowly as the temperature difference between the VPM and the air evolves \citep{harrington2018thesis}. 
On shorter time scales, this can be thought of as baseline offset.}. 
To minimize its impact on the single scan data, we leveraged its asymmetry between the orthogonal detector pairs and used the pair-add demodulated data \citep[equivalent to pair-differencing for the unmodulated experiment,][Cleary et al., in prep.]{harrington21} for the rest of the analysis. 
We keep the notations $u/v$ for the pair-add data for simplicity.
The demodulation process depends on modeling of the VPM transfer function, and its error could result in a bias in the polarization amplitude or cause leakage between linear and circular polarization. 
One of the main factors in the model is the assumed sky spectrum (the cloud linear polarization and atmospheric circular polarization) due to the frequency dependence of the VPM modulation efficiency within the detector bandpass. 
CLASS data were demodulated assuming a sky spectrum consisting of CMB and Galactic foreground contribution in linear polarization and atmospheric emission in circular polarization \citep{petroff20}. 
The assumed linear polarization spectrum is shallower than the fiducial spectrum from Rayleigh scattering from clouds, which inevitably leads to polarization bias. 
We denote the bias to linear polarization and linear-to-circular polarization leakage from VPM transfer function error by $B_{uu}$ and $B_{vu}$, respectively. 
The modeling of these quantities requires accurate calibration of the VPM parameters; however, in the absence of such calibration, we elect to empirically determine these bias factors from the data in Section \ref{ssec:circular-bias}.

The demodulated data can be related to the sky polarization signal\footnote{Throughout this work, we distinguish the polarization signals in the detector frame and on the sky by small and capitalized letters.} $[Q, U, V]$ through
\begin{align}
    \begin{bmatrix}
        u\\ v 
    \end{bmatrix} &=
    \begin{bmatrix}
        -\sin2(\gamma+\phi_P) & \cos2(\gamma+\phi_P) & 0\\
        0 & 0 & 1
    \end{bmatrix}
    \begin{bmatrix}
        Q\\ U\\ V
    \end{bmatrix},
    \label{eq:pointing-matrix}
\end{align}
where $\gamma$ is the detector position angle on the sky and $\phi_P$ is a small angle correction accounting for the VPM wire direction as viewed by each detector \citep{Li23}.
The sky maps were solved by inverse-variance weighting the demodulated data from all detectors:
\begin{align}
    \begin{bmatrix}
        Q\\
        U\\
        V\\ 
    \end{bmatrix}=
    \left(\mathbf{P}^T\mathbf{N}^{-1}\mathbf{P}\right)^{-1}\mathbf{P}^T\mathbf{N}^{-1}
    \begin{bmatrix}
        u\\
        v\\
    \end{bmatrix},
    \label{eq:map-making}
\end{align}
where $\mathbf{P}$ is the pointing matrix in Equation \ref{eq:pointing-matrix} and $\mathbf{N}$ is the diagonal noise matrix. 
To guard against the bright cloud signal biasing the variance estimates, each detector time stream was first outlier-removed using an iterative median absolute deviation (MAD) algorithm where for each iteration samples with absolute deviation from the median greater than five times the median of the absolute deviation from the median were flagged as outliers. 
A singular-value decomposition was then performed on the outlier-removed data to isolate the common modes. 
The per-span variance of each detector was estimated after projecting out modes with singular values four times greater than the median.
Prior to the map-making, the median of the outlier-removed data was subtracted from each detector to prevent the residual VPM synchronous emission or other long-time-scale drifts from biasing the result.

Limited by the instantaneous detector coverage of the transient cloud events, we binned the data into azimuthal maps (hereafter \textit{az-maps}), where close-by data in the time-azimuth coordinates are averaged over regardless of the elevation pointing.
For all frequencies, the azimuth resolution was set to $1^\circ$, and temporal resolution was chosen such that each continuous $360^\circ$ scan was binned into the same temporal bin. Depending on the scanning speed, the temporal resolution is around $180$--$360~\mathrm{s}$.
The az-maps trade off spatial resolution in the elevation direction for polarization sensitivity; therefore, elevation structures of the cloud smaller than the focal plane size are lost. 
Examples of the multifrequency Stokes $Q/U$ az-maps are shown in Figure \ref{fig:multi-freq-az-maps}.

\subsection{Cloud Detection}\label{ssec:detection}
The search for cloud signals in CLASS data aims to find any burst-like linear polarization signal that has temporal and azimuthal contiguity without assuming a priori knowledge of the cloud polarization properties.
Instead of operating on the \emph{sky} az-maps introduced in Equation \ref{eq:map-making} for cloud detection, we made separate az-maps for the detector frame $u$ signal by mapping it as an intensity signal, i.e., disregarding the position angle dependency in Equation \ref{eq:map-making}. 
After the map-making as outlined in Section \ref{sec:class-polarimetry}, we filtered out from the $u$ az-map a low-order polynomial component in the temporal direction with outlier rejection based on MAD. 
The polynomial order was chosen to be the integer number of hours of each span. 

Depending on the boresight angle of the telescope, the $-Q$ polarization from the cloud can appear as bursts in the $u$-az-map with positive (positive boresight) or negative (negative boresight) signs. 
The search was conducted for both cases, where pixel values exceeding seven times the map white noise level (propagated from the $u$ time stream variances) were flagged.
Assuming the smoothness of the cloud signal, the boundaries of flagged pixels were further expanded and tailored using consecutive morphological closing and opening algorithms. These morphological operations take into account the periodic boundary condition in the azimuth direction.
Each isolated contiguous flagged region was identified as a cloud candidate, and an additional morphological dilation operation was applied to expand the region for background estimations.
The sky polarization signals were extracted from the sky az-maps for each Stokes parameter at the corresponding region with background subtraction. 
The variance in the background region was assigned as the measurement uncertainty.

\begin{figure*}
\begin{center}
\includegraphics{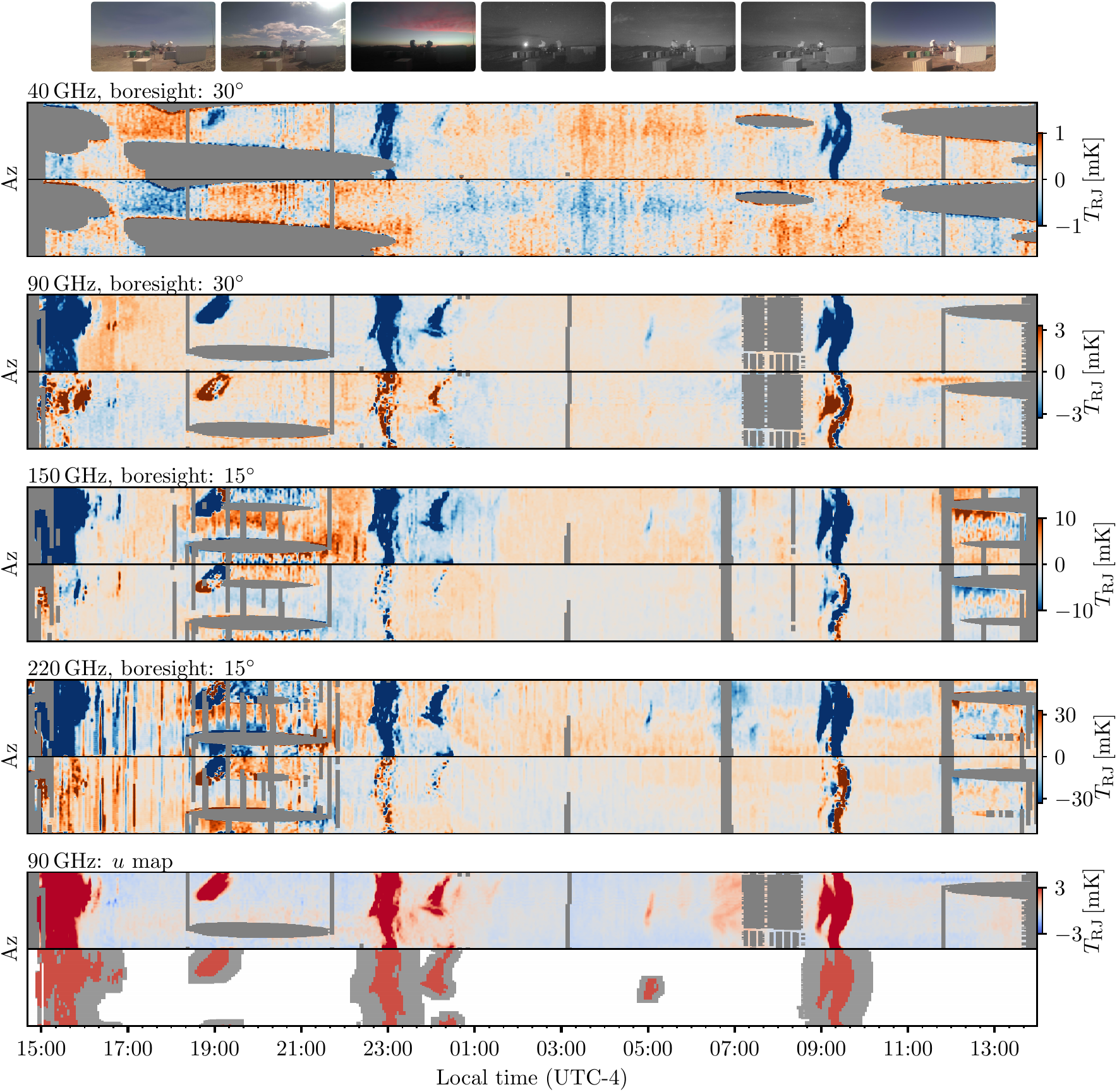}
\end{center}
\caption{
    Example of multifrequency polarized cloud observation starting on 2022-01-15.
    \textit{Top:} Sky images recorded by camera 3, shown in 3-hour intervals. As noted in Figure \ref{fig:camera-pano}, the highest elevation of camera 3 is below the lowest elevation of the telescope; therefore, clouds in the camera image do not directly correspond to the az-map detections. 
    \textit{Row 1--8:} Stokes $Q$ (top) and $U$ (bottom) az-maps for four CLASS frequency bands. 
    The $x$-axis represents time, and the $y$-axis, from bottom to top of each panel, is the $-180$ to $180^\circ$ azimuth angle.
    The azimuthal coverage of camera 3 roughly corresponds to the lower quarter of the az-maps.
    The gray area shows missing data due to Sun avoidance or excessive detector flux jumps.
    \textit{Row 9:} az-map of the detector-frame Stokes $u$ map for the $90~\mathrm{GHz}$ band.
    \textit{Row 10:} the cloud candidates identified from the $90~\mathrm{GHz}$ $u$ az-map. Red patches show the identified cloud candidates and the gray patches show the extended background region.
    All az-maps and camera images have the temporal axes aligned.  
    \label{fig:multi-freq-az-maps}
}
\end{figure*}

As an example, the last two panels of Figure \ref{fig:multi-freq-az-maps} show the $90~\mathrm{GHz}$ $u$-az-map and the cloud detection flags for the span between 2022-01-25 and 2022-01-26. 
For the population analysis below, the same extraction procedure was performed independently for each of the four CLASS bands for every span.
In total, we have identified 1481, 3677, 4110, and 4178 cloud candidates at 40, 90, 150, and 220 GHz, respectively.

\subsection{Circular Polarization and Demodulation Bias}\label{ssec:circular-bias}

\begin{deluxetable}{ccccc}
    \tablecaption{CLASS demodulation bias \label{tab:demod-bias}}
    \tablehead{\colhead{Band} & \colhead{40\tablenotemark{a}} & \colhead{90\tablenotemark{b}} &\colhead{150}& \colhead{220}}
    \startdata
    \specialrule{.05em}{0em}{0em}
    $\hat{B}_{uu}$ & 1.080/1.078 & 0.983/0.935 & 0.966 & 0.943  \\
    $\hat{B}_{vu}$ & $-0.215$/$-0.186$ & $-0.124$/$-0.255$ & $-0.200$ & $-0.257$ \\
    \specialrule{.05em}{0em}{0em}
    \enddata
    \tablenotetext{a}{The estimated bias terms are given separately for data before and after 2020-01-20. } 
    \tablenotetext{b}{The estimated bias terms are given separately for data before and after 2019-05-20. } 
\end{deluxetable}

\begin{figure}
    \begin{center}
    \includegraphics{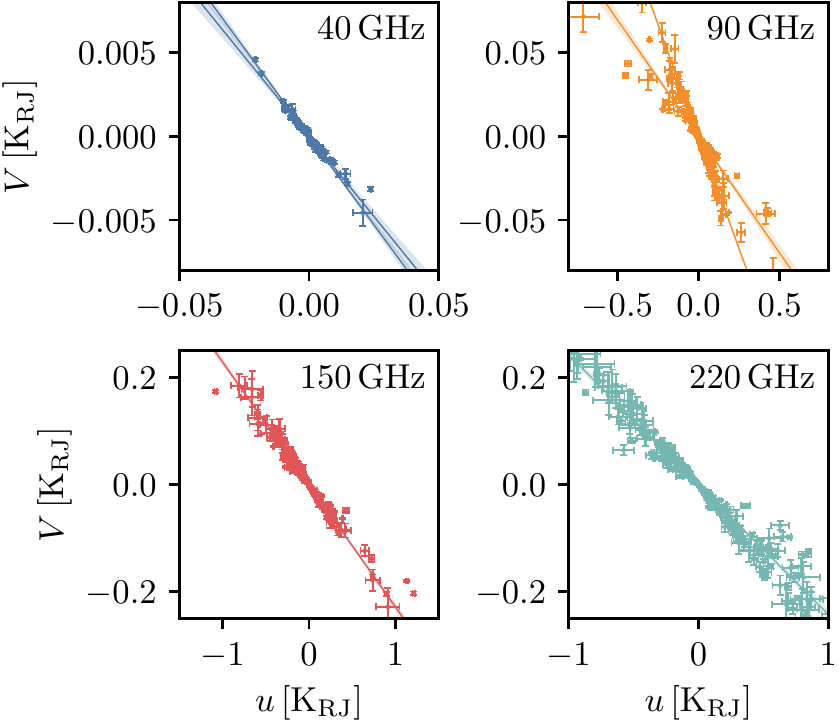}
    \end{center}
    \caption{Detected circular polarization as a function of the Stokes $u$ amplitude.
    The colored line and the shaded region indicate the best linear fit between the two variables and should be interpreted as the measured transfer function bias term $\hat{B}_{vu}$ and its uncertainty.
    Separate fits are performed for the 40~GHz data before and after 2021-02-11 due to the replacement of the VPM grid \citep{Li23}.
    The $90~\mathrm{GHz}$ panel also suggests a separate fit for data before and after 2019-05-20 due to a bimodal transition in the VPM system.
    \label{fig:circular-leakage}
    }
\end{figure}

Clouds are not expected to produce circular polarization, and the reflection mechanisms outlined in Section \ref{sec:physics} should not yield significant circular polarization due to the low Rayleigh scattering optical depth.
However, non-zero circular polarization signals were found in az-maps at all frequencies. 
Figure \ref{fig:circular-leakage} shows the measured circular polarization as a function of the linear polarization amplitude measured in the detector frame ($u$) for samples with a $\mathrm{S/N}>5$.
Since circular polarization is correlated with the detector-frame quantity instead of the sky signal ($Q/U$), the majority of the signal must be the result of a polarization leakage presumably due to inaccurate VPM transfer function modeling.
Considering the uncertainties of the measurement in both directions, we use orthogonal distance regression to fit a scaling factor for each frequency.
The results are empirical estimates of the aforementioned demodulation bias ${B}_{vu}$ for cloud polarization. These numbers are summarized in Table \ref{tab:demod-bias} and denoted as $\hat{B}_{vu}$. 
For $90~\mathrm{GHz}$ we found distinct trends before and after 2019-05-20, which is likely from a perturbation to the VPM system during telescope maintenance.
By simulating the modulation and demodulation process with various VPM parameters and sky spectra, we found tight linear correlations between $B_{vu}$ and $B_{uu}$; therefore, the $\hat{B}_{vu}$ measurements can be used for $B_{uu}$ estimates in the absence of precise VPM parameter determination. 
These $\hat{B}_{uu}$ estimates are also summarized in Table \ref{tab:demod-bias}, and used to correct for the linear polarization bias in the following analysis.

The significant linear polarization and its unique spectral shape make clouds promising targets (among other sources with different spectral dependencies) for VPM transfer function calibration \citep{Li23}. 
This possibility will be explored and applied to the cosmological analysis for future CLASS data.

\section{Cloud Population Analysis}\label{sec:population}
\subsection{Confirmation with Camera Images}
\begin{figure}
    \begin{center}
    \includegraphics{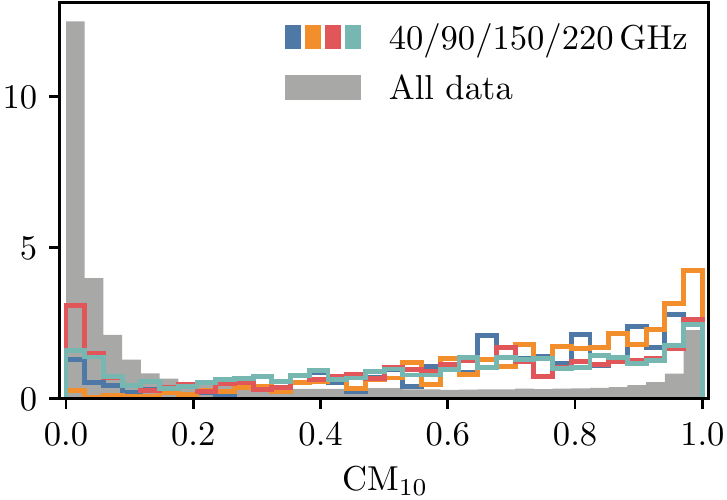}
    \end{center}
    \caption{
        The distribution of cloud measure above $10^\circ$ elevation ($\mathrm{CM}_{10}$) associated with each of the CLASS polarization detections.
        The colored histograms correspond to the polarization detections from each of the four frequencies. 
        The global $\mathrm{CM}_{10}$ distribution from all data is shown in gray.
    \label{fig:cm-confirmation}
    }
\end{figure}

In addition to the visual confirmation from individual examples (e.g., Figure \ref{fig:multi-freq-az-maps}), we compare the radiometer detection to the cloud measure from the site cameras.
Figure \ref{fig:cm-confirmation} shows the distribution of the cloud measure ($\mathrm{CM}_{10}$) associated with CLASS radiometer detections. 
The average $\mathrm{CM}_{10}$ values within a 15-minute window around each of the cloud candidates were extracted.
For detections at all four frequencies, the associated higher-than-global cloud measures indicate a significant correspondence between the radiometer detection and the clouds optically identified by the cameras.
The single-sided Mann--Whitney U test suggests that the $\mathrm{CM}_{10}$ during the time period with radiometer polarization detection is significantly higher than the global distribution to the at least $22.7\sigma$ level (for the lowest $40~\mathrm{GHz}$ detections).
Due to the aforementioned FoV mismatch between the site camera and the telescopes and the finite cadence of the camera images, the significance shown here should be considered as a lower limit.

\subsection{Polarization Angle}
\begin{figure}[h]
    \begin{center}
    \includegraphics{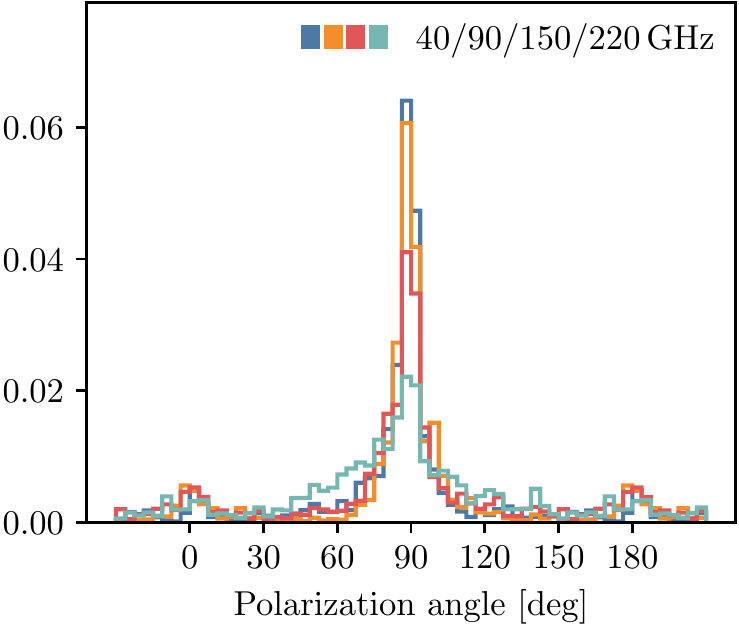}
    \end{center}
    \caption{
        Polarization angle distribution of the cloud candidates from each of the four CLASS frequency bands.
        The histograms are weighted by the observation time of each cloud event.
        The boundaries of the histograms are periodically extended by $30^\circ$ to better show the features around $0^\circ$ ($180^\circ$).
    \label{fig:population-pol-angle}
    }
\end{figure}

The polarization angle for each candidate was estimated as
\begin{equation}
    \psi = \frac{1}{2}\arctan\left(\frac{\bar U}{\bar Q}\right),
\end{equation}
where $\bar U$ and $\bar Q$ are the background-subtracted averaged Stokes amplitudes in the flagged region (Section \ref{ssec:detection}).
For boresight angles $\pm45^\circ$ and $0^\circ$, either $Q$ or $U$ may have large uncertainties due to the degeneracy between the cloud polarization direction and the detector angle, and the simple estimate here would have large uncertainty and/or bias. 
Therefore, we avoid the interpretation of polarization angles from individual measurements and only study the cloud population as a whole.
Figure \ref{fig:population-pol-angle} shows the polarization angle distribution of each cloud candidate weighted by the event observation time. 
It is worth noting that the seven boresight rotations of the telescopes have different sensitivities to cloud polarization. 
At $\pm45^\circ$ boresight, the VPM and detectors have the most sensitivity to the $Q$ polarization, but the uncertainty on the Stokes $U$ component is the greatest; therefore, the polarization angle cannot be reliably determined. 
We find the average polarization angle to be $85.7\pm 1.5^\circ$, $91.1\pm 0.5^\circ$, $89.6\pm 0.5^\circ$, and $86.4\pm 0.6^\circ$ for the $40$, $90$, $150$, and $220~\mathrm{GHz}$ samples, respectively.
Despite the unfavorable conditions for the polarization angle measurement, the distribution is consistent with the predicted $90^\circ$ polarization angle from the scattering of a population of horizontally aligned cloud particles.
The distributions in Figure \ref{fig:population-pol-angle} also show a slight preference for angles around $0^\circ$ ($180^\circ$). 
By visually inspecting the camera images, we found that they are associated with clouds at very low altitudes close to the ground (but not exclusively), and the polarization angle may be explained by $+Q$ polarization from the scattering of the atmosphere radiation by spherical cloud particles like water droplets (Section~\ref{sec:physics}).

\subsection{Polarization Spectrum}\label{ssec:spectra}
CLASS's multifrequency cloud observations provide a unique opportunity to study the spectrum of the cloud polarization and to verify the polarization mechanisms as outlined in Section~\ref{sec:physics}.
\begin{figure}
    \begin{center}
    \includegraphics{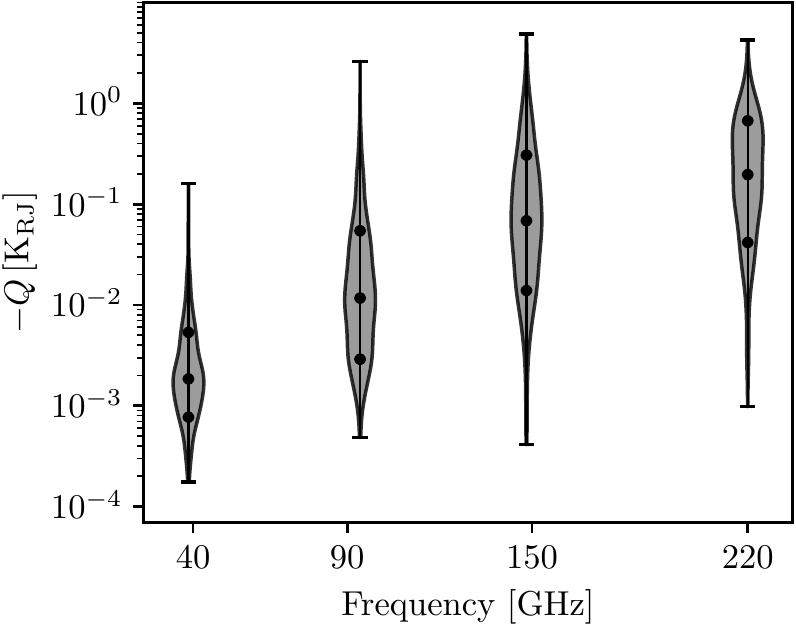}
    \end{center}
    \caption{Measured cloud polarization signal ($-Q$) as a function of the effective frequency. 
    The samples include all $90~\mathrm{GHz}$-selected candidates with $\mathrm{S/N}>4$ at each band. 
    The dots mark for each distribution the $16$, $50$, and $84$ percentiles. 
    \label{fig:population-spectrum}
    }
\end{figure}

In order to compare the multifrequency data, we chose the candidates detected from the $90~\mathrm{GHz}$ data and performed forced photometry on the other three bands. 
The forced photometry finds corresponding pixels in other bands by matching the temporal and azimuthal bins, which might not always be possible due to the different pointing directions of the telescope mounts and/or the missing data.
To account for the bias resulting from the VPM transfer function error, $\hat{B}_{uu}$ was divided out from the polarization measurements from each band.
The power measured by the CLASS bolometers also needs to be corrected for the absorption by the water vapor. This is especially relevant for cloud observations that typically occur during high PWV conditions. 
However, we lack dedicated PWV measurement to accurately determine the real-time PWV, and resort to the PWV values reported by the nearby APEX weather station at 1-minute cadence. 
For each cloud candidate, we queried the maximum PWV reported by APEX within a 10-minute window and calculated the atmospheric transmission based on the detector bandpass \citep{appel22}.

Figure \ref{fig:population-spectrum} summarizes the polarization amplitudes ($-Q$) from the forced photometry for all the candidates with S/N greater than four.
The spectral index $\alpha$ can be obtained by jointly fitting the polarization amplitudes across two or more frequency bands. 
Using the 375 samples with detections at $90$ and $150~\mathrm{GHz}$ with $\mathrm{S/N}>4$, we find the index $\alpha=3.90\pm0.06$, consistent with the Rayleigh scattering model of ice particles as outlined in Section \ref{sec:physics}.
However, when the $40$ or $220~\mathrm{GHz}$ samples are included, the best-fit $\alpha$ parameters significantly deviate from the nominal value of $4$.
These results suggest excess/deficit of power at $40/200~\mathrm{GHz}$, which we discuss below.

\subsubsection{Effects From Liquid Water}
At temperatures above $-40~\mathrm{^\circ C}$, liquid water may coexist with ice crystals in its supercooled state \citep{korolev2022}. 
The dielectric permittivity of liquid water within the microwave range has a strong frequency dependency, therefore the Rayleigh scattering spectrum of liquid water has an additional term $|(\epsilon_\nu-1)/(\epsilon_\nu+2)|^2$ (Equation \ref{eq:tau-formula}).
However, the liquid water fraction is expected to be low in high clouds, moreover, the spherical shape and smallness of the water droplets \citep[about $20~\mathrm{\mu m}$,][]{Lasher-Trapp2005} also make liquid water scattering a subdominant contributor to the observed polarization signals.
Their absorption effect, however, could be important since the absorption cross-section, in the Rayleigh regime, is proportional to $D^3$ (whereas the scattering is proportional to $D^6$, where $D$ is the particle diameter) and the liquid water absorption efficiency is a few orders-of-magnitude higher than that of ice due to its large imaginary part of the dielectric permittivity \citep{turner2016improved}.
Considering a toy model with a layer of liquid water at the base of the ice cloud, the liquid water absorption can be incorporated by inserting into the integral of equation \ref{eq:scattering-formula}:
\begin{equation}
\exp{\left[-k_m(\epsilon_{\nu, T})\mathrm{LWP}\left(\frac{1}{|\cos\theta|}+\frac{1}{|\cos\delta|}\right)\right]},
\end{equation}
where $k_m$ is the mass absorption coefficient computed from the complex permittivity that also depends on the frequency and temperature $T$; LWP is the column mass density of the liquid water layer. The two secant terms correspond to the absorption of the incident and scattered light, respectively.
The nadir angle ($\theta$) dependency of the absorption adjusts the scattering contribution from different directions, therefore, the final result also depends on the particle shape/orientation in the ice cloud. 

\begin{figure}
    \begin{center}
    \includegraphics{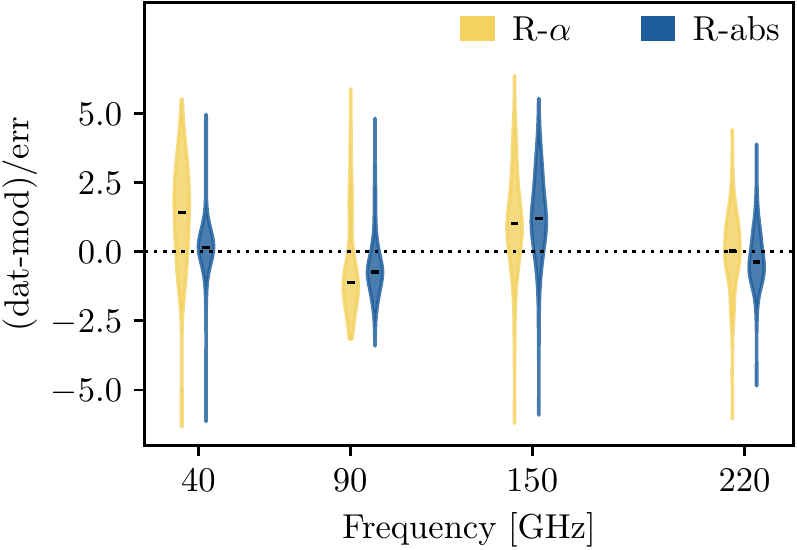}
    \end{center}
    \caption{Best-fit residual distribution for the spectral models in comparison.
    The yellow and blue colors represent R-$\alpha$, and R-abs, respectively. Each component of the graph shows the distribution of the residuals from the best-fit model. 
    The two distributions for each frequency are arbitrarily offset for visualization.
    \label{fig:spec-chi}
    }
\end{figure}

We compared two models of the polarization spectrum below: R-$\alpha$ is a power-law model that has the spectral index $\alpha$ as a free parameter; R-abs includes the liquid water layer absorption and assumes the Rayleigh scattering spectrum ($\alpha=4$).
We jointly fit these models to 131 samples with joint detection at all four frequency bands with $\mathrm{S/N}>4$, and the residuals are displayed in Figure \ref{fig:spec-chi}.
As mentioned in the section above, a purely Rayleigh scattering spectrum is not a good fit across all four frequencies. Specifically, the R-$\alpha$ model yields a best-fit $\alpha=3.17\pm0.05$. The excess of power at $40~\mathrm{GHz}$ and the favor of a lower spectral index are evident in Figure \ref{fig:spec-chi}, and the mean $\chi^2$ per data degree-of-freedom is $3.7$. 
By including additional liquid absorption, model R-abs brings more consistency between $40$ and $90~\mathrm{GHz}$, as indicated by the improved normality of the distributions and the lower mean $\chi^2=2.3$. 
We note that the absorption model depends on the cloud temperature, LWP, and particle shape (through $\Delta_z$ in Equation \ref{eq:permittivity}), and the regression problem is underdetermined with the four-frequency photometry. 
Therefore, the result here only demonstrates the possibility that the presence of liquid water could play a role in the spectrum of cloud polarization, but the interpretation of the best-fit parameters should be conducted with caution.

\subsubsection{Spectra at 220 GHz}
\begin{figure}
    \begin{center}
    \includegraphics{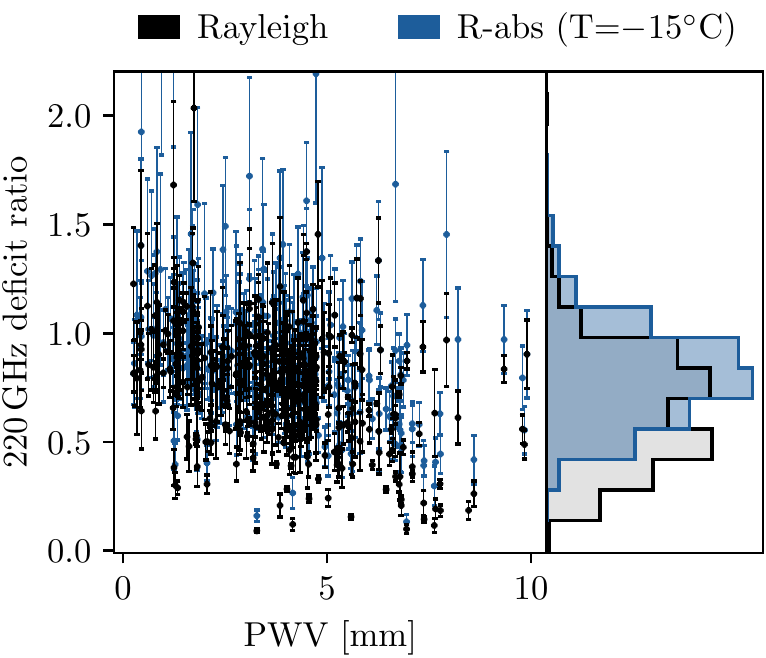}
    \end{center}
    \caption{Polarization amplitude ratio between the $220~\mathrm{GHz}$ and $150~\mathrm{GHz}$ scaled by the Rayleigh scattering spectrum.
    The histograms on the right panel are weighted by the observation time of each cloud event.
    The black points correspond to the Rayleigh model, i.e., the $220/150~\mathrm{GHz}$ ratios are divided by $(220/148.2)^{4}$. 
    The blue points additionally account for the liquid water absorption, where the absorption term is fit jointly with $90$,$150$, and $220~\mathrm{GHz}$ data for each sample assuming the R-abs model at cloud temperature $-15~\mathrm{^\circ C}$.
    \label{fig:population-220-deficit}
    }
\end{figure}

To assess the potential deviation from Rayleigh scattering at around $220~\mathrm{GHz}$, we selected 375 samples with joint detection at $150/220~\mathrm{GHz}$ with S/N greater than four. 
Figure~\ref{fig:population-220-deficit} plots in black the ratio of the Stokes $Q$ signal measured at $220~\mathrm{GHz}$ compared to the $150~\mathrm{GHz}$ measurement scaled by $(220/148.2)^{4}$, assuming the Rayleigh scattering spectrum.
The left panel shows the deficit ratio against PWV, and the right panel is the histogram weighted by the cloud observation time, centering around the mean value $0.61\pm0.02$.
A weak correlation with PWV is found with Pearson coefficient $-0.33\pm0.03$; however, it is unlikely that the deficit is due to the unaccounted-for water vapor absorption. 
The additional $40\%$ absorption would require PWV values greater than $20~\mathrm{mm}$, which is improbable at the CLASS site.

Nonetheless, as discussed in the previous section, it is possible that supercooled liquid water is contributing to the absorption.
We jointly fit the $90$, $150$, and $220~\mathrm{GHz}$ data (315 samples with $\mathrm{S/N}>4$) with model R-abs, but fixed the cloud temperature to $-15~\mathrm{^\circ C}$ to reduce the number of free parameters. 
The result of the $220~\mathrm{GHz}$ deficit ratio after correcting for the absorption with the best-fit R-abs parameters are shown in Figure \ref{fig:population-220-deficit} in blue.
The best-fit model can explain most of the deficit with a wide spread of LWP values -- the $50$, $84$, and $95\%$ quantiles are $0.4$, $1.2$, and $2.0~\mathrm{kg\,m^{-2}}$, respectively.
These values are much higher than the typical value in the Chajnantor area \citep[below $0.01~\mathrm{kg~\mathrm{m}^{-2}}$ for $90\%$ of the time,][]{Kuo2017} based on MERRA-2 or the maximum value $0.6~\mathrm{kg\,\mathrm{m}^{-2}}$ found in the ERA5 database over the survey periods. 
However, these records are based on reanalyses and are averaged values over a wide area of the atmosphere, and might not be applicable to individual cloud measurements. 
Lacking targeted LWP measurement for each sample, we cannot preclude the possibility that the clouds are in mixed-phase and the liquid water absorption contributes (at least partially) to the deviation of the Rayleigh scattering spectrum at $220~\mathrm{GHz}$.

Alternatively, the apparent deficit can be interpreted as evidence of a population of large ice particles that contribute to polarization through resonant scattering.
As is shown in Figure \ref{fig:mie-spectrum}, the polarization amplitude is sharply suppressed at around $220~\mathrm{GHz}$ for particle diameter $\gtrsim 350~\mathrm{\mu m}$. 
Considering a population of cloud particles, depending on size distribution \citep[which primarily depends on the temperature,][]{Austin+2009}, the deficit ratios range between $20\sim80\%$  for temperature $-50\sim-80\mathrm{^\circ C}$.
We note that the calculation above assumed Mie scattering from spherical particles, and the exact deficit ratio might be sensitive to the size, shape, and orientation of the cloud particles, which is beyond the scope of this work.

\section{Conclusion}\label{sec:conclusion}
We presented multifrequency observations of cloud polarization from CLASS at frequencies centered around 40, 90, 150, and 220~GHz. 
The az-mapping technique identifies transient sky polarization signals and is less susceptible to time-domain glitches than searches in time-ordered data.
A blind search for localized polarization transient signals in daytime and nighttime throughout the 2016--2022 observation period yielded significant detection of cloud events that concur with the cloud images captured by the site cameras and correlate with the cloud measure extracted from the cloud images based on a CNN classifier. 

Leveraging the polarization modulator, finite on-sky position angle offsets of the detectors at the $45^\circ$ elevation, and the az-maps, we were able to combine the data from multiple detectors to perform polarization measurement of the transient signal from clouds on a single scan.
At the threshold chosen for cloud detection, we identified 1481, 3677, 4110, and 4178 cloud candidates at 40, 90, 150, and 220~GHz, respectively. 
The polarization angle distribution of the majority of cloud events is consistent with the predicted $90^\circ$ polarization angle from the scattering of a population of horizontally aligned ice crystals.
At a lower significance, we also found a group of events with vertical polarizations, which may be attributed to the polarization effect of spherical cloud particles (e.g., liquid water).

Focusing on the cloud candidates selected by the $90~\mathrm{GHz}$ data, we compared the $-Q$ polarization signal at all four frequencies. 
The joint spectrum fit from $90/150~\mathrm{GHz}$ data shows a mean spectral index $\alpha=3.90\pm0.06$ consistent with the ice crystal Rayleigh scattering. 
The polarization amplitudes at $40/220~\mathrm{GHz}$ show excess/deficit compared to the $\alpha=4$ spectrum.
Using a toy model with a layer of supercooled liquid water at the base of the ice cloud, we found better agreement between the data and Rayleigh spectra with liquid water absorption. 
Although the fitting result is underdetermined due to the finite frequency sampling of CLASS data and lack of targeted LWP measurement, the result hints that the presence of water might be important to explain the spectral shape of cloud polarization, but the exact solution would depend upon the modeling of cloud particle size/shape distribution, their orientation, the ice/liquid water mixing, and cloud temperature.
Alternatively, the $60\%$ deficit ratio observed at $220~\mathrm{GHz}$ measurement compared to the scaled $150~\mathrm{GHz}$ data assuming $\alpha=4$ can be explained by a population of cloud particles with diameters greater than $350~\mathrm{\mu m}$ that contribute to the polarization through the Mie scattering. 
This picture is also consistent with the long-tail distribution of the ice crystal size in high clouds \citep{Austin+2009} within the temperature range $-50\sim-80\mathrm{^\circ C}$.

No circular polarization signal was detected from the clouds. All apparent circular polarization signals are consistent with a linear-to-circular polarization leakage from VPM modeling uncertainty. 
Leveraging the large amplitude of the cloud signal, this leakage signal will be used in future CLASS analysis to constrain the VPM parameters for the cosmological dataset.

As a final remark, the cloud events analyzed in this work were deliberately selected from the brightest end of the population. 
For the cosmological analysis, the time-ordered data around these events are likely flagged in the pipeline either due to the anomalously large signal or coincident bad weather and/or excessive detector flux jumps.
The fainter end of the cloud population, on the other hand, is more difficult to avoid with data selection and could impact the long-term stability of the instrument as a polarization ``noise" \citep[][Cleary et al. in prep.]{harrington21}
Since 2022, we have installed an all-sky camera to better monitor the cloud conditions and to guide the survey.
The methods developed and lessons learned here will be used to improve the CMB data reduction for CLASS, and are available for future experiments \citep{SO,S4} that aim to push the limit of CMB observations further from the ground.

\section{Acknowledgments}
Figure \ref{fig:cartoon} has been designed using assets from Freepik.com.
We acknowledge primary funding support for CLASS from the National Science Foundation Division of Astronomical Sciences under Grant Numbers 0959349, 1429236, 1636634, 1654494, 2034400, and 2109311. We thank Johns Hopkins University President R. Daniels and the Deans of the Kreiger School of Arts and Sciences for their steadfast support of CLASS. We further acknowledge the very generous support of Jim and Heather Murren (JHU A\&S ’88), Matthew Polk (JHU A\&S Physics BS ’71), David Nicholson, and Michael Bloomberg (JHU Engineering ’64). The CLASS project employs detector technology developed in collaboration between JHU and Goddard Space Flight Center under several previous and ongoing NASA grants. Detector development work at JHU was funded by NASA cooperative agreement 80NSSC19M0005. CLASS is located in the Parque Astron\'omico Atacama in northern Chile under the auspices of the Agencia Nacional de Investigaci\'on y Desarrollo (ANID). 
We acknowledge scientific and engineering contributions from 
Max Abitbol, 
Aamir Ali,
Fletcher Boone,
Michael~K. Brewer,
Sarah~Marie Bruno,
David Carcamo, 
Carol Chan, 
Manwei Chan, 
Joseph~Cleary,
Kevin~L. Denis,
Francisco Espinoza, 
Benjamín Fernández, 
Pedro Flux\'a Rojas,
Joey Golec, 
Dominik Gothe, 
Ted Grunberg, 
Mark Halpern, 
Saianeesh Haridas, 
Kyle Helson, 
Gene Hilton, 
Connor Henley, 
Johannes Hubmayr,
John Karakla,
Lindsay Lowry, 
Jeffrey~John McMahon, 
Nick Mehrle, 
Nathan J.~Miller,
Carolina~Morales Perez, 
Carolina N\'{u}\~{n}ez,
Keisuke Osumi,
Ivan~L. Padilla, 
Gonzalo Palma, 
Lucas Parker, 
Sasha Novack, 
Bastian Pradenas, Isu Ravi, 
Carl~D. Reintsema, 
Gary Rhoades, Daniel Swartz, Bingjie Wang, Qinan Wang, 
Tiffany Wei, 
Janet~L. Weiland,
Zi\'ang Yan,
Lingzhen Zeng
and Zhuo Zhang. For essential logistical support, we thank Jill Hanson, William Deysher, Joseph Zolenas, LaVera Jackson, Miguel Angel D\'iaz, Mar\'ia Jos\'e Amaral, and Chantal Boisvert. 
We acknowledge productive collaboration with the JHU Physical Sciences Machine Shop team.
R. D\"unner thanks ANID for grant BASAL CATA FB210003.
Z.X. is supported by the Gordon and Betty Moore Foundation through grant GBMF5215 to the Massachusetts Institute of Technology.

\software{
    {\tt numpy} \citep{numpy20}, 
    {\tt scipy} \citep{scipy}, 
    {\tt astropy} \citep{astropy}, 
    {\tt matplotlib} \citep{matplotlib}, 
    {\tt PyTorch} \citep{pytorch},
    {\tt flask} \citep{flask},
    {\tt am} \citep{am}, 
    {\tt miepython} \citep{miepython},
    {\tt astrometry.net} \citep{astrometrynet}.
}

\appendix
\section{CNN-based Cloud classification}
\label{sec:cnn}
Convolutional neural network (CNN) classifiers are powerful for image classification and have been demonstrated to be effective for cloud monitoring in all-sky images \citep{Mommert2020}.
We adopted the pre-trained \texttt{AlexNet} model \citep{alexnet} and fine-tuned the network for our specific task.

The site camera images were selected across a wide range of season/observation conditions and UTC hours and were cropped into $240\times240$-pixel patches for evaluation. 
Each patch is classified to be either \textit{clear}, \textit{cloud}, or \textit{obscured} (by the ground, the telescope, or celestial objects that prohibit a definitive cloud identification), disregarding the type, thickness, and coverage of cloud within the patch.
An internal web application was set up to collect manual classifications from the CLASS collaboration. 
A total of 19,095 valid classifications were recorded throughout this project.

Each classification was based solely on the pixels within the patch, which is approximately a few degrees on the sky and is sufficient in most cases; however, there are cases where single patch evaluation would fail. 
For example, pixels around the Sun and near the horizon during twilight appear white due to the Mie scattering of the aerosol and are hard to distinguish from clouds in a single patch (the annual modulation around twilight in Figure~\ref{fig:cloud-annual}). 
Although the model is capable of also distinguishing the morphology of the cloud \citep{Zhang+2018}, we only attempt dichotomous labeling due to the lower quality images taken at night time, and no finer classification was made regarding the volume or thickness of the cloud. 
A continuous cloudiness measurement was achieved by taking the average of the binary evaluation of all patches in the image, but this is often an overestimation due to the sparsity (e.g., altocumulus) or diffuseness (e.g., cirrus) of the cloud. 
The nighttime images add extra complexities as the cameras occasionally experience after-images that last for a few minutes, which triggers false detection and degrades the temporal resolution of the measurements.

\begin{figure}[h!]
\centering
\includegraphics[width=0.76\linewidth]{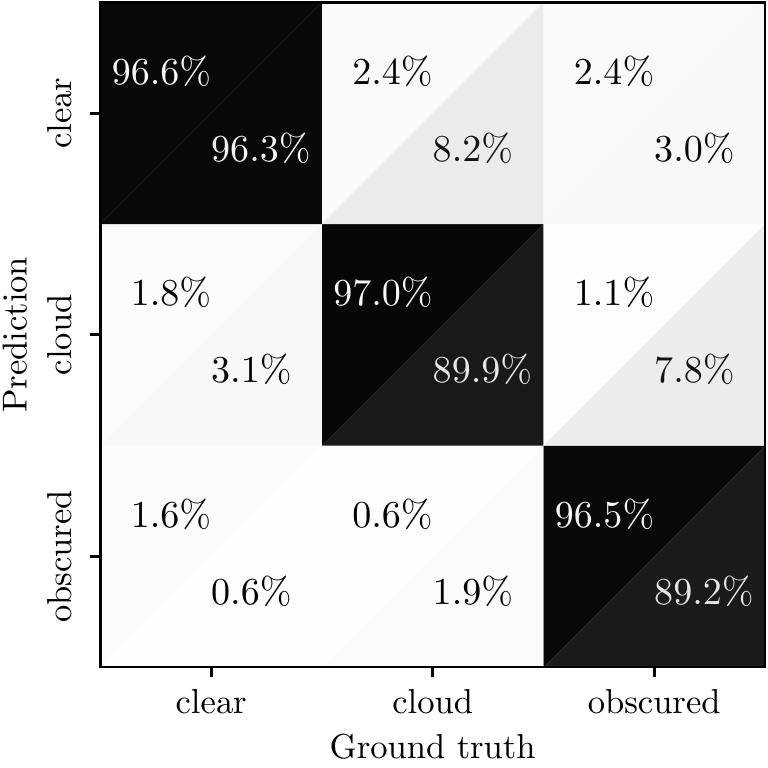}
\caption{The confusion matrix of the CNN classifier. The numbers in the diagonal (off-diagonal) blocks are the true-positive (true-negative) rate for each column. The two values in each block correspond to the statistics for daytime (upper-left) and nighttime (lower-right).}
\label{fig:confusion_matrix}
\end{figure}

Using 9548 labeled patches as the training set, the classifier achieved an overall $\gtrsim 96\%$ ($\gtrsim 89\%$) accuracy for cloud detection at daytime (nighttime) on the rest 9547 testing samples.
The full confusion matrix of the classifier is shown in Figure~\ref{fig:confusion_matrix}. 
The about $10\%$ false-negative rate for cloud classification at night is almost the limit for manual classification imposed by the challenges addressed above. 

\bibliography{class_common, software, software_common, main, cmb}

\begin{thebibliography}{77}
\expandafter\ifx\csname natexlab\endcsname\relax\def\natexlab#1{#1}\fi

\bibitem[{{Abazajian} {et~al.}(2016){Abazajian}, {Adshead}, {Ahmed}, {Allen},
  {Alonso}, {Arnold}, {Baccigalupi}, {Bartlett}, {Battaglia}, {Benson},
  {Bischoff}, {Borrill}, {Buza}, {Calabrese}, {Caldwell}, {Carlstrom}, {Chang},
  {Crawford}, {Cyr-Racine}, {De Bernardis}, {de Haan}, {di Serego Alighieri},
  {Dunkley}, {Dvorkin}, {Errard}, {Fabbian}, {Feeney}, {Ferraro}, {Filippini},
  {Flauger}, {Fuller}, {Gluscevic}, {Green}, {Grin}, {Grohs}, {Henning},
  {Hill}, {Hlozek}, {Holder}, {Holzapfel}, {Hu}, {Huffenberger}, {Keskitalo},
  {Knox}, {Kosowsky}, {Kovac}, {Kovetz}, {Kuo}, {Kusaka}, {Le Jeune}, {Lee},
  {Lilley}, {Loverde}, {Madhavacheril}, {Mantz}, {Marsh}, {McMahon},
  {Meerburg}, {Meyers}, {Miller}, {Munoz}, {Nguyen}, {Niemack}, {Peloso},
  {Peloton}, {Pogosian}, {Pryke}, {Raveri}, {Reichardt}, {Rocha}, {Rotti},
  {Schaan}, {Schmittfull}, {Scott}, {Sehgal}, {Shandera}, {Sherwin}, {Smith},
  {Sorbo}, {Starkman}, {Story}, {van Engelen}, {Vieira}, {Watson}, {Whitehorn},
  \& {Kimmy Wu}}]{S4}
{Abazajian}, K.~N., {Adshead}, P., {Ahmed}, Z., {et~al.} 2016,
  \href{https://ui.adsabs.harvard.edu/abs/2016arXiv161002743A}{\href{http://dx.doi.org/10.48550/arXiv.1610.02743}{\color{magenta}arXiv
  e-prints}, arXiv:1610.02743}

\bibitem[{{Ade} {et~al.}(2019){Ade}, {Aguirre}, {Ahmed}, {Aiola}, {Ali},
  {Alonso}, {Alvarez}, {Arnold}, {Ashton}, {Austermann}, \& et~al.}]{SO}
{Ade}, P., {Aguirre}, J., {Ahmed}, Z., {et~al.} 2019,
  \href{http://dx.doi.org/10.1088/1475-7516/2019/02/056}{\color{magenta}\jcap},
  \href{https://ui.adsabs.harvard.edu/abs/2019JCAP...02..056A}{2019, 056}

\bibitem[{{Aiola} {et~al.}(2020){Aiola}, {Calabrese}, {Maurin}, {Naess},
  {Schmitt}, {Abitbol}, {Addison}, {Ade}, {Alonso}, {Amiri}, {Amodeo},
  {Angile}, {Austermann}, {Baildon}, {Battaglia}, {Beall}, {Bean}, {Becker},
  {Bond}, {Bruno}, {Calafut}, {Campusano}, {Carrero}, {Chesmore}, {Cho},
  {Choi}, {Clark}, {Cothard}, {Crichton}, {Crowley}, {Darwish}, {Datta},
  {Denison}, {Devlin}, {Duell}, {Duff}, {Duivenvoorden}, {Dunkley},
  {D{\"u}nner}, {Essinger-Hileman}, {Fankhanel}, {Ferraro}, {Fox}, {Fuzia},
  {Gallardo}, {Gluscevic}, {Golec}, {Grace}, {Gralla}, {Guan}, {Hall},
  {Halpern}, {Han}, {Hargrave}, {Hasselfield}, {Helton}, {Henderson},
  {Hensley}, {Hill}, {Hilton}, {Hilton}, {Hincks}, {Hlo{\v{z}}ek}, {Ho},
  {Hubmayr}, {Huffenberger}, {Hughes}, {Infante}, {Irwin}, {Jackson}, {Klein},
  {Knowles}, {Koopman}, {Kosowsky}, {Lakey}, {Li}, {Li}, {Li}, {Lokken},
  {Louis}, {Lungu}, {MacInnis}, {Madhavacheril}, {Maldonado}, {Mallaby-Kay},
  {Marsden}, {McMahon}, {Menanteau}, {Moodley}, {Morton}, {Namikawa}, {Nati},
  {Newburgh}, {Nibarger}, {Nicola}, {Niemack}, {Nolta}, {Orlowski-Sherer},
  {Page}, {Pappas}, {Partridge}, {Phakathi}, {Pisano}, {Prince}, {Puddu}, {Qu},
  {Rivera}, {Robertson}, {Rojas}, {Salatino}, {Schaan}, {Schillaci}, {Sehgal},
  {Sherwin}, {Sierra}, {Sievers}, {Sifon}, {Sikhosana}, {Simon}, {Spergel},
  {Staggs}, {Stevens}, {Storer}, {Sunder}, {Switzer}, {Thorne}, {Thornton},
  {Trac}, {Treu}, {Tucker}, {Vale}, {Van Engelen}, {Van Lanen}, {Vavagiakis},
  {Wagoner}, {Wang}, {Ward}, {Wollack}, {Xu}, {Zago}, \& {Zhu}}]{aiola20}
{Aiola}, S., {Calabrese}, E., {Maurin}, L., {et~al.} 2020,
  \href{http://dx.doi.org/10.1088/1475-7516/2020/12/047}{\color{magenta}\jcap},
  \href{https://ui.adsabs.harvard.edu/abs/2020JCAP...12..047A}{2020, 047}

\bibitem[{{Appel} {et~al.}(2022){Appel}, {Bennett}, {Brewer}, {Bustos}, {Chan},
  {Chuss}, {Cleary}, {Couto}, {Dahal}, {Datta}, {Denis}, {Eimer},
  {Essinger-Hileman}, {Harrington}, {Iuliano}, {Li}, {Marriage},
  {N{\'u}{\~n}ez}, {Osumi}, {Padilla}, {Petroff}, {Rostem}, {Valle}, {Watts},
  {Weiland}, {Wollack}, \& {Xu}}]{appel22}
{Appel}, J.~W., {Bennett}, C.~L., {Brewer}, M.~K., {et~al.} 2022,
  \href{http://dx.doi.org/10.3847/1538-4365/ac8cf2}{\color{magenta}\apjs},
  \href{https://ui.adsabs.harvard.edu/abs/2022ApJS..262...52A}{262, 52}

\bibitem[{{Astropy Collaboration} {et~al.}(2022){Astropy Collaboration},
  {Price-Whelan}, {Lim}, {Earl}, {Starkman}, {Bradley}, {Shupe}, {Patil},
  {Corrales}, {Brasseur}, {N{\"o}the}, {Donath}, {Tollerud}, {Morris},
  {Ginsburg}, {Vaher}, {Weaver}, {Tocknell}, {Jamieson}, {van Kerkwijk},
  {Robitaille}, {Merry}, {Bachetti}, {G{\"u}nther}, {Aldcroft},
  {Alvarado-Montes}, {Archibald}, {B{\'o}di}, {Bapat}, {Barentsen},
  {Baz{\'a}n}, {Biswas}, {Boquien}, {Burke}, {Cara}, {Cara}, {Conroy},
  {Conseil}, {Craig}, {Cross}, {Cruz}, {D'Eugenio}, {Dencheva}, {Devillepoix},
  {Dietrich}, {Eigenbrot}, {Erben}, {Ferreira}, {Foreman-Mackey}, {Fox},
  {Freij}, {Garg}, {Geda}, {Glattly}, {Gondhalekar}, {Gordon}, {Grant},
  {Greenfield}, {Groener}, {Guest}, {Gurovich}, {Handberg}, {Hart},
  {Hatfield-Dodds}, {Homeier}, {Hosseinzadeh}, {Jenness}, {Jones}, {Joseph},
  {Kalmbach}, {Karamehmetoglu}, {Ka{\l}uszy{\'n}ski}, {Kelley}, {Kern},
  {Kerzendorf}, {Koch}, {Kulumani}, {Lee}, {Ly}, {Ma}, {MacBride}, {Maljaars},
  {Muna}, {Murphy}, {Norman}, {O'Steen}, {Oman}, {Pacifici}, {Pascual},
  {Pascual-Granado}, {Patil}, {Perren}, {Pickering}, {Rastogi}, {Roulston},
  {Ryan}, {Rykoff}, {Sabater}, {Sakurikar}, {Salgado}, {Sanghi}, {Saunders},
  {Savchenko}, {Schwardt}, {Seifert-Eckert}, {Shih}, {Jain}, {Shukla}, {Sick},
  {Simpson}, {Singanamalla}, {Singer}, {Singhal}, {Sinha}, {Sip{\H{o}}cz},
  {Spitler}, {Stansby}, {Streicher}, {{\v{S}}umak}, {Swinbank}, {Taranu},
  {Tewary}, {Tremblay}, {Val-Borro}, {Van Kooten}, {Vasovi{\'c}}, {Verma}, {de
  Miranda Cardoso}, {Williams}, {Wilson}, {Winkel}, {Wood-Vasey}, {Xue},
  {Yoachim}, {Zhang}, {Zonca}, \& {Astropy Project Contributors}}]{astropy}
{Astropy Collaboration}, {Price-Whelan}, A.~M., {Lim}, P.~L., {et~al.} 2022,
  \href{http://dx.doi.org/10.3847/1538-4357/ac7c74}{\color{magenta}\apj},
  \href{https://ui.adsabs.harvard.edu/abs/2022ApJ...935..167A}{935, 167}

\bibitem[{{Austin} {et~al.}(2009){Austin}, {Heymsfield}, \&
  {Stephens}}]{Austin+2009}
{Austin}, R.~T., {Heymsfield}, A.~J., \& {Stephens}, G.~L. 2009,
  \href{http://dx.doi.org/10.1029/2008JD010049}{\color{magenta}Journal of
  Geophysical Research (Atmospheres)},
  \href{https://ui.adsabs.harvard.edu/abs/2009JGRD..114.0A23A}{114, D00A23}

\bibitem[{{Battistelli} {et~al.}(2012){Battistelli}, {Amico}, {Ba{\`u}},
  {Berg{\'e}}, {Br{\'e}elle}, {Charlassier}, {Collin}, {Cruciani}, {de
  Bernardis}, {Dufour}, {Dumoulin}, {Gervasi}, {Giard}, {Giordano},
  {Giraud-H{\'e}raud}, {Guglielmi}, {Hamilton}, {Land{\'e}}, {Maffei},
  {Maiello}, {Marnieros}, {Masi}, {Passerini}, {Piacentini}, {Piat},
  {Piccirillo}, {Pisano}, {Polenta}, {Rosset}, {Salatino}, {Schillaci},
  {Sordini}, {Spinelli}, {Tartari}, \& {Zannoni}}]{battistelli2012}
{Battistelli}, E.~S., {Amico}, G., {Ba{\`u}}, A., {et~al.} 2012,
  \href{http://dx.doi.org/10.1111/j.1365-2966.2012.20951.x}{\color{magenta}\mnras},
  \href{https://ui.adsabs.harvard.edu/abs/2012MNRAS.423.1293B}{423, 1293}

\bibitem[{{BICEP/Keck Collaboration} {et~al.}(2022){BICEP/Keck Collaboration},
  {Ade}, {Ahmed}, {Amiri}, {Barkats}, {Thakur}, {Bischoff}, {Beck}, {Bock},
  {Boenish}, {Bullock}, {Buza}, {Cheshire}, {Connors}, {Cornelison},
  {Crumrine}, {Cukierman}, {Denison}, {Dierickx}, {Duband}, {Eiben},
  {Fatigoni}, {Filippini}, {Fliescher}, {Goeckner-Wald}, {Goldfinger},
  {Grayson}, {Grimes}, {Hall}, {Halal}, {Halpern}, {Hand}, {Harrison},
  {Henderson}, {Hildebrandt}, {Hilton}, {Hubmayr}, {Hui}, {Irwin}, {Kang},
  {Karkare}, {Karpel}, {Kefeli}, {Kernasovskiy}, {Kovac}, {Kuo}, {Lau},
  {Leitch}, {Lennox}, {Megerian}, {Minutolo}, {Moncelsi}, {Nakato}, {Namikawa},
  {Nguyen}, {O'Brient}, {Ogburn}, {Palladino}, {Prouve}, {Pryke}, {Racine},
  {Reintsema}, {Richter}, {Schillaci}, {Schwarz}, {Schmitt}, {Sheehy},
  {Soliman}, {Germaine}, {Steinbach}, {Sudiwala}, {Teply}, {Thompson}, {Tolan},
  {Tucker}, {Turner}, {Umilt{\`a}}, {Verg{\`e}s}, {Vieregg}, {Wandui}, {Weber},
  {Wiebe}, {Willmert}, {Wong}, {Wu}, {Yang}, {Yoon}, {Young}, {Yu}, {Zeng},
  {Zhang}, \& {Zhang}}]{BK-XV20}
{BICEP/Keck Collaboration}, {Ade}, P.~A.~R., {Ahmed}, Z., {et~al.} 2022,
  \href{http://dx.doi.org/10.3847/1538-4357/ac4886}{\color{magenta}\apj},
  \href{https://ui.adsabs.harvard.edu/abs/2022ApJ...927...77A}{927, 77}

\bibitem[{{Bohren} \& {Huffman}(1983)}]{Bohren&Huffman1983}
{Bohren}, C.~F. \& {Huffman}, D.~R. 1983, {Absorption and scattering of light
  by small particles} (John Wiley \& Sons, Ltd)

\bibitem[{{Br{\'e}on} \& {Dubrulle}(2004)}]{breon2004}
{Br{\'e}on}, F.-M. \& {Dubrulle}, B. 2004,
  \href{http://dx.doi.org/10.1175/JAS-3309.1}{\color{magenta}Journal of the
  Atmospheric Sciences},
  \href{https://ui.adsabs.harvard.edu/abs/2004JAtS...61.2888B}{61, 2888}

\bibitem[{{Chepfer}(1999)}]{chepfer1999}
{Chepfer}, H. 1999,
  \href{http://dx.doi.org/10.1016/S0022-4073(99)00036-9}{\color{magenta}\jqsrt},
  \href{https://ui.adsabs.harvard.edu/abs/1999JQSRT..63..521C}{63, 521}

\bibitem[{{Chuss} {et~al.}(2012){Chuss}, {Wollack}, {Henry}, {Hui}, {Juarez},
  {Krejny}, {Moseley}, \& {Novak}}]{chus12vpm}
{Chuss}, D.~T., {Wollack}, E.~J., {Henry}, R., {et~al.} 2012,
  \href{http://dx.doi.org/10.1364/AO.51.000197}{\color{magenta}\ao},
  \href{http://adsabs.harvard.edu/abs/2012ApOpt..51..197C}{51, 197}

\bibitem[{{Cort{\'e}s} {et~al.}(2020){Cort{\'e}s}, {Cort{\'e}s}, {Reeves},
  {Bustos}, \& {Radford}}]{Cortes2020}
{Cort{\'e}s}, F., {Cort{\'e}s}, K., {Reeves}, R., {Bustos}, R., \& {Radford},
  S. 2020,
  \href{http://dx.doi.org/10.1051/0004-6361/202037784}{\color{magenta}\aap},
  \href{https://ui.adsabs.harvard.edu/abs/2020A&A...640A.126C}{640, A126}

\bibitem[{{Dahal} {et~al.}(2022){Dahal}, {Appel}, {Datta}, {Brewer}, {Ali},
  {Bennett}, {Bustos}, {Chan}, {Chuss}, {Cleary}, {Couto}, {Denis},
  {D{\"u}nner}, {Eimer}, {Espinoza}, {Essinger-Hileman}, {Golec}, {Harrington},
  {Helson}, {Iuliano}, {Karakla}, {Li}, {Marriage}, {McMahon}, {Miller},
  {Novack}, {N{\'u}{\~n}ez}, {Osumi}, {Padilla}, {Palma}, {Parker}, {Petroff},
  {Reeves}, {Rhoades}, {Rostem}, {Valle}, {Watts}, {Weiland}, {Wollack}, \&
  {Xu}}]{dahal22}
{Dahal}, S., {Appel}, J.~W., {Datta}, R., {et~al.} 2022,
  \href{http://dx.doi.org/10.3847/1538-4357/ac397c}{\color{magenta}\apj},
  \href{https://ui.adsabs.harvard.edu/abs/2022ApJ...926...33D}{926, 33}

\bibitem[{{Davis} {et~al.}(2005){Davis}, {Wu}, {Emde}, {Jiang}, {Cofield}, \&
  {Harwood}}]{Davis2005}
{Davis}, C.~P., {Wu}, D.~L., {Emde}, C., {et~al.} 2005,
  \href{http://dx.doi.org/10.1029/2005GL022681}{\color{magenta}\grl},
  \href{https://ui.adsabs.harvard.edu/abs/2005GeoRL..3214806D}{32, L14806}

\bibitem[{{Davis} {et~al.}(2007){Davis}, {Avallone}, {Weinstock}, {Twohy},
  {Smith}, \& {Kok}}]{davis2007comparisons}
{Davis}, S.~M., {Avallone}, L.~M., {Weinstock}, E.~M., {et~al.} 2007,
  \href{http://dx.doi.org/10.1029/2006JD008214}{\color{magenta}Journal of
  Geophysical Research (Atmospheres)},
  \href{https://ui.adsabs.harvard.edu/abs/2007JGRD..11210212D}{112, D10212}

\bibitem[{{Defer} {et~al.}(2014){Defer}, {Galligani}, {Prigent}, \&
  {Jimenez}}]{Defer+2014}
{Defer}, E., {Galligani}, V.~S., {Prigent}, C., \& {Jimenez}, C. 2014,
  \href{http://dx.doi.org/10.1002/2014JD022353}{\color{magenta}Journal of
  Geophysical Research (Atmospheres)},
  \href{https://ui.adsabs.harvard.edu/abs/2014JGRD..11912301D}{119, 12,301}

\bibitem[{{Dutcher} {et~al.}(2021){Dutcher}, {Balkenhol}, {Ade}, {Ahmed},
  {Anderes}, {Anderson}, {Archipley}, {Avva}, {Aylor}, {Barry}, {Basu Thakur},
  {Benabed}, {Bender}, {Benson}, {Bianchini}, {Bleem}, {Bouchet}, {Bryant},
  {Byrum}, {Carlstrom}, {Carter}, {Cecil}, {Chang}, {Chaubal}, {Chen}, {Cho},
  {Chou}, {Cliche}, {Crawford}, {Cukierman}, {Daley}, {de Haan}, {Denison},
  {Dibert}, {Ding}, {Dobbs}, {Everett}, {Feng}, {Ferguson}, {Foster}, {Fu},
  {Galli}, {Gambrel}, {Gardner}, {Goeckner-Wald}, {Gualtieri}, {Guns}, {Gupta},
  {Guyser}, {Halverson}, {Harke-Hosemann}, {Harrington}, {Henning}, {Hilton},
  {Hivon}, {Holder}, {Holzapfel}, {Hood}, {Howe}, {Huang}, {Irwin}, {Jeong},
  {Jonas}, {Jones}, {Khaire}, {Knox}, {Kofman}, {Korman}, {Kubik}, {Kuhlmann},
  {Kuo}, {Lee}, {Leitch}, {Lowitz}, {Lu}, {Meyer}, {Michalik}, {Millea},
  {Montgomery}, {Nadolski}, {Natoli}, {Nguyen}, {Noble}, {Novosad}, {Omori},
  {Padin}, {Pan}, {Paschos}, {Pearson}, {Posada}, {Prabhu}, {Quan},
  {Raghunathan}, {Rahlin}, {Reichardt}, {Riebel}, {Riedel}, {Rouble}, {Ruhl},
  {Sayre}, {Schiappucci}, {Shirokoff}, {Smecher}, {Sobrin}, {Stark}, {Stephen},
  {Story}, {Suzuki}, {Thompson}, {Thorne}, {Tucker}, {Umilta}, {Vale},
  {Vanderlinde}, {Vieira}, {Wang}, {Whitehorn}, {Wu}, {Yefremenko}, {Yoon},
  {Young}, \& {SPT-3G Collaboration}}]{SPT3G-2021}
{Dutcher}, D., {Balkenhol}, L., {Ade}, P.~A.~R., {et~al.} 2021,
  \href{http://dx.doi.org/10.1103/PhysRevD.104.022003}{\color{magenta}\prd},
  \href{https://ui.adsabs.harvard.edu/abs/2021PhRvD.104b2003D}{104, 022003}

\bibitem[{{Eastman} \& {Warren}(2014)}]{Eastman2014}
{Eastman}, R. \& {Warren}, S.~G. 2014,
  \href{http://dx.doi.org/10.1175/JCLI-D-13-00352.1}{\color{magenta}Journal of
  Climate}, \href{https://ui.adsabs.harvard.edu/abs/2014JCli...27.2386E}{27,
  2386}

\bibitem[{{Eimer} {et~al.}(2023){Eimer}, {Li}, {Brewer}, {Shi}, {Ali}, {Appel},
  {Bennett}, {Bustos}, {Chuss}, {Cleary}, {Dahal}, {Datta}, {Denes Couto},
  {Denis}, {D{\"u}nner}, {Essinger-Hileman}, {Flux{\'a}}, {Hubmayer},
  {Harrington}, {Iuliano}, {Karakla}, {Marriage}, {N{\'u}{\~n}ez}, {Parker},
  {Petroff}, {Reeves}, {Rostem}, {Valle}, {Watts}, {Weiland}, {Wollack}, {Xu},
  \& {Zeng}}]{eimer23}
{Eimer}, J.~R., {Li}, Y., {Brewer}, M.~K., {et~al.} 2023,
  \href{https://ui.adsabs.harvard.edu/abs/2023arXiv230900675E}{\href{http://dx.doi.org/10.48550/arXiv.2309.00675}{\color{magenta}arXiv
  e-prints}, arXiv:2309.00675}

\bibitem[{{Ellison}(2007)}]{Ellison2007}
{Ellison}, W.~J. 2007,
  \href{http://dx.doi.org/10.1063/1.2360986}{\color{magenta}Journal of Physical
  and Chemical Reference Data},
  \href{https://ui.adsabs.harvard.edu/abs/2007JPCRD..36....1E}{36, 1}

\bibitem[{{Errard} {et~al.}(2015){Errard}, {Ade}, {Akiba}, {Arnold}, {Atlas},
  {Baccigalupi}, {Barron}, {Boettger}, {Borrill}, {Chapman}, {Chinone},
  {Cukierman}, {Delabrouille}, {Dobbs}, {Ducout}, {Elleflot}, {Fabbian},
  {Feng}, {Feeney}, {Gilbert}, {Goeckner-Wald}, {Halverson}, {Hasegawa},
  {Hattori}, {Hazumi}, {Hill}, {Holzapfel}, {Hori}, {Inoue}, {Jaehnig},
  {Jaffe}, {Jeong}, {Katayama}, {Kaufman}, {Keating}, {Kermish}, {Keskitalo},
  {Kisner}, {Le Jeune}, {Lee}, {Leitch}, {Leon}, {Linder}, {Matsuda},
  {Matsumura}, {Miller}, {Myers}, {Navaroli}, {Nishino}, {Okamura}, {Paar},
  {Peloton}, {Poletti}, {Puglisi}, {Rebeiz}, {Reichardt}, {Richards}, {Ross},
  {Rotermund}, {Schenck}, {Sherwin}, {Siritanasak}, {Smecher}, {Stebor},
  {Steinbach}, {Stompor}, {Suzuki}, {Tajima}, {Takakura}, {Tikhomirov},
  {Tomaru}, {Whitehorn}, {Wilson}, {Yadav}, \& {Zahn}}]{Errard2015}
{Errard}, J., {Ade}, P.~A.~R., {Akiba}, Y., {et~al.} 2015,
  \href{http://dx.doi.org/10.1088/0004-637X/809/1/63}{\color{magenta}\apj},
  \href{https://ui.adsabs.harvard.edu/abs/2015ApJ...809...63E}{809, 63}

\bibitem[{{Essinger-Hileman} {et~al.}(2014){Essinger-Hileman}, {Ali}, {Amiri},
  {Appel}, {Araujo}, {Bennett}, {Boone}, {Chan}, {Cho}, {Chuss}, {Colazo},
  {Crowe}, {Denis}, {D{\"u}nner}, {Eimer}, {Gothe}, {Halpern}, {Harrington},
  {Hilton}, {Hinshaw}, {Huang}, {Irwin}, {Jones}, {Karakla}, {Kogut}, {Larson},
  {Limon}, {Lowry}, {Marriage}, {Mehrle}, {Miller}, {Miller}, {Moseley},
  {Novak}, {Reintsema}, {Rostem}, {Stevenson}, {Towner}, {U-Yen}, {Wagner},
  {Watts}, {Wollack}, {Xu}, \& {Zeng}}]{essinger-hileman14spie}
{Essinger-Hileman}, T., {Ali}, A., {Amiri}, M., {et~al.} 2014,
  \href{http://dx.doi.org/10.1117/12.2056701}{\color{magenta}\procspie},
  \href{https://ui.adsabs.harvard.edu/abs/2014SPIE.9153E..1IE}{9153, 91531I}

\bibitem[{{Feofilov} \& {Stubenrauch}(2019)}]{Feofilov2019}
{Feofilov}, A.~G. \& {Stubenrauch}, C.~J. 2019,
  \href{http://dx.doi.org/10.5194/acp-19-13957-2019}{\color{magenta}Atmospheric
  Chemistry \& Physics},
  \href{https://ui.adsabs.harvard.edu/abs/2019ACP....1913957F}{19, 13957}

\bibitem[{{Gong} \& {Wu}(2017)}]{Gong&Wu2017}
{Gong}, J. \& {Wu}, D.~L. 2017,
  \href{http://dx.doi.org/10.5194/acp-17-2741-2017}{\color{magenta}Atmospheric
  Chemistry \& Physics},
  \href{https://ui.adsabs.harvard.edu/abs/2017ACP....17.2741G}{17, 2741}

\bibitem[{Grinberg(2018)}]{flask}
Grinberg, M. 2018, Flask web development: developing web applications with
  python (" O'Reilly Media, Inc.")

\bibitem[{{Harrington} {et~al.}(2021){Harrington}, {Datta}, {Osumi}, {Ali},
  {Appel}, {Bennett}, {Brewer}, {Bustos}, {Chan}, {Chuss}, {Cleary}, {Denes
  Couto}, {Dahal}, {D{\"u}nner}, {Eimer}, {Essinger-Hileman}, {Hubmayr}, {Raul
  Espinoza Inostroza}, {Iuliano}, {Karakla}, {Li}, {Marriage}, {Miller},
  {N{\'u}{\~n}ez}, {Padilla}, {Parker}, {Petroff}, {Pradenas M{\'a}rquez},
  {Reeves}, {Flux{\'a} Rojas}, {Rostem}, {Augusto Nunes Valle}, {Watts},
  {Weiland}, {Wollack}, {Xu}, \& {Class Collaboration}}]{harrington21}
{Harrington}, K., {Datta}, R., {Osumi}, K., {et~al.} 2021,
  \href{http://dx.doi.org/10.3847/1538-4357/ac2235}{\color{magenta}\apj},
  \href{https://ui.adsabs.harvard.edu/abs/2021ApJ...922..212H}{922, 212}

\bibitem[{{Harrington} {et~al.}(2018){Harrington}, {Eimer}, {Chuss}, {Petroff},
  {Cleary}, {DeGeorge}, {Grunberg}, {Ali}, {Appel}, {Bennett}, {Brewer},
  {Bustos}, {Chan}, {Couto}, {Dahal}, {Denis}, {D{\"u}nner},
  {Essinger-Hileman}, {Fluxa}, {Halpern}, {Hilton}, {Hinshaw}, {Hubmayr},
  {Iuliano}, {Karakla}, {Marriage}, {McMahon}, {Miller}, {Nu{\~n}ez},
  {Padilla}, {Palma}, {Parker}, {Pradenas Marquez}, {Reeves}, {Reintsema},
  {Rostem}, {Augusto Nunes Valle}, {Van Engelhoven}, {Wang}, {Wang}, {Watts},
  {Weiland}, {Wollack}, {Xu}, {Yan}, \& {Zeng}}]{harrington18spie}
{Harrington}, K., {Eimer}, J., {Chuss}, D.~T., {et~al.} 2018,
  \href{http://dx.doi.org/10.1117/12.2313614}{\color{magenta}\procspie},
  \href{http://adsabs.harvard.edu/abs/2018SPIE10708E..2MH}{10708, 107082M}

\bibitem[{{Harrington}(2018)}]{harrington2018thesis}
{Harrington}, K. M.~K. 2018,
  \href{https://ui.adsabs.harvard.edu/abs/2018PhDT.......173H}{{Variable-delay
  polarization modulators for the CLASS telescopes}}, PhD thesis, Johns Hopkins
  University, Maryland

\bibitem[{Harris {et~al.}(2020)Harris, Millman, van~der Walt, Gommers,
  Virtanen, Cournapeau, Wieser, Taylor, Berg, Smith, Kern, Picus, Hoyer, van
  Kerkwijk, Brett, Haldane, del R{\'{i}}o, Wiebe, Peterson,
  G{\'{e}}rard-Marchant, Sheppard, Reddy, Weckesser, Abbasi, Gohlke, \&
  Oliphant}]{numpy20}
Harris, C.~R., Millman, K.~J., van~der Walt, S.~J., {et~al.} 2020,
  \href{http://dx.doi.org/10.1038/s41586-020-2649-2}{\color{magenta}Nature},
  585, 357

\bibitem[{{Hashino} {et~al.}(2014){Hashino}, {Chiruta}, {Polzin}, {Kubicek}, \&
  {Wang}}]{Hashino2014}
{Hashino}, T., {Chiruta}, M., {Polzin}, D., {Kubicek}, A., \& {Wang}, P.~K.
  2014,
  \href{http://dx.doi.org/10.1016/j.atmosres.2014.07.003}{\color{magenta}Atmospheric
  Research}, \href{https://ui.adsabs.harvard.edu/abs/2014AtmRe.150...79H}{150,
  79}

\bibitem[{{Hendry} \& {McCormick}(1976)}]{Hendry&McCormick1976}
{Hendry}, A. \& {McCormick}, G.~C. 1976,
  \href{http://dx.doi.org/10.1029/JC081i030p05353}{\color{magenta}\jgr},
  \href{https://ui.adsabs.harvard.edu/abs/1976JGR....81.5353H}{81, 5353}

\bibitem[{{Hersbach} {et~al.}(2020){Hersbach}, {Bell}, {Berrisford},
  {Hirahara}, {Hor{\'a}nyi}, {Mu{\~n}oz-Sabater}, {Nicolas}, {Peubey}, {Radu},
  {Schepers}, {Simmons}, {Soci}, {Abdalla}, {Abellan}, {Balsamo}, {Bechtold},
  {Biavati}, {Bidlot}, {Bonavita}, {Chiara}, {Dahlgren}, {Dee}, {Diamantakis},
  {Dragani}, {Flemming}, {Forbes}, {Fuentes}, {Geer}, {Haimberger}, {Healy},
  {Hogan}, {H{\'o}lm}, {Janiskov{\'a}}, {Keeley}, {Laloyaux}, {Lopez}, {Lupu},
  {Radnoti}, {Rosnay}, {Rozum}, {Vamborg}, {Villaume}, \& {Th{\'e}paut}}]{ERA5}
{Hersbach}, H., {Bell}, B., {Berrisford}, P., {et~al.} 2020,
  \href{http://dx.doi.org/10.1002/qj.3803}{\color{magenta}Quarterly Journal of
  the Royal Meteorological Society},
  \href{https://ui.adsabs.harvard.edu/abs/2020QJRMS.146.1999H}{146, 1999}

\bibitem[{{Heymsfield} {et~al.}(2002){Heymsfield}, {Bansemer}, {Field},
  {Durden}, {Stith}, {Dye}, {Hall}, \& {Grainger}}]{Heymsfield+2002}
{Heymsfield}, A.~J., {Bansemer}, A., {Field}, P.~R., {et~al.} 2002,
  \href{http://dx.doi.org/10.1175/1520-0469(2002)059<3457:OAPOPS>2.0.CO;2}{\color{magenta}Journal
  of Atmospheric Sciences},
  \href{https://ui.adsabs.harvard.edu/abs/2002JAtS...59.3457H}{59, 3457}

\bibitem[{{Hunter}(2007)}]{matplotlib}
{Hunter}, J.~D. 2007,
  \href{http://dx.doi.org/10.1109/MCSE.2007.55}{\color{magenta}Computing in
  Science and Engineering},
  \href{https://ui.adsabs.harvard.edu/abs/2007CSE.....9...90H}{9, 90}

\bibitem[{{Johnson} {et~al.}(2007){Johnson}, {Collins}, {Abroe}, {Ade}, {Bock},
  {Borrill}, {Boscaleri}, {de Bernardis}, {Hanany}, {Jaffe}, {Jones}, {Lee},
  {Levinson}, {Matsumura}, {Rabii}, {Renbarger}, {Richards}, {Smoot},
  {Stompor}, {Tran}, {Winant}, {Wu}, \& {Zuntz}}]{Johnson07-HWP}
{Johnson}, B.~R., {Collins}, J., {Abroe}, M.~E., {et~al.} 2007,
  \href{http://dx.doi.org/10.1086/518105}{\color{magenta}\apj},
  \href{https://ui.adsabs.harvard.edu/abs/2007ApJ...665...42J}{665, 42}

\bibitem[{{Kawata}(1978)}]{kawata1978}
{Kawata}, Y. 1978,
  \href{http://dx.doi.org/10.1016/0019-1035(78)90035-0}{\color{magenta}\icarus},
  \href{https://ui.adsabs.harvard.edu/abs/1978Icar...33..217K}{33, 217}

\bibitem[{{King} \& {Lubin}(2016)}]{King&Lubin2016}
{King}, S. \& {Lubin}, P. 2016,
  \href{http://dx.doi.org/10.1103/PhysRevD.94.023501}{\color{magenta}\prd},
  \href{https://ui.adsabs.harvard.edu/abs/2016PhRvD..94b3501K}{94, 023501}

\bibitem[{{Klett}(1995)}]{klett1995orientation}
{Klett}, J.~D. 1995,
  \href{http://dx.doi.org/10.1175/1520-0469(1995)052<2276:OMFPIT>2.0.CO;2}{\color{magenta}Journal
  of Atmospheric Sciences},
  \href{https://ui.adsabs.harvard.edu/abs/1995JAtS...52.2276K}{52, 2276}

\bibitem[{{Korolev} \& {Milbrandt}(2022)}]{korolev2022}
{Korolev}, A. \& {Milbrandt}, J. 2022,
  \href{http://dx.doi.org/10.1029/2022GL099578}{\color{magenta}\grl},
  \href{https://ui.adsabs.harvard.edu/abs/2022GeoRL..4999578K}{49, e99578}

\bibitem[{Krizhevsky {et~al.}(2012)Krizhevsky, Sutskever, \& Hinton}]{alexnet}
Krizhevsky, A., Sutskever, I., \& Hinton, G.~E. 2012, 25

\bibitem[{{Kuo}(2017)}]{Kuo2017}
{Kuo}, C.-L. 2017,
  \href{http://dx.doi.org/10.3847/1538-4357/aa8b74}{\color{magenta}\apj},
  \href{https://ui.adsabs.harvard.edu/abs/2017ApJ...848...64K}{848, 64}

\bibitem[{{Lang} {et~al.}(2010){Lang}, {Hogg}, {Mierle}, {Blanton}, \&
  {Roweis}}]{astrometrynet}
{Lang}, D., {Hogg}, D.~W., {Mierle}, K., {Blanton}, M., \& {Roweis}, S. 2010,
  \href{http://dx.doi.org/10.1088/0004-6256/139/5/1782}{\color{magenta}\aj},
  \href{https://ui.adsabs.harvard.edu/abs/2010AJ....139.1782L}{139, 1782}

\bibitem[{{Lasher-Trapp} {et~al.}(2005){Lasher-Trapp}, {Cooper}, \&
  {Blyth}}]{Lasher-Trapp2005}
{Lasher-Trapp}, S.~G., {Cooper}, W.~A., \& {Blyth}, A.~M. 2005,
  \href{http://dx.doi.org/10.1256/qj.03.199}{\color{magenta}Quarterly Journal
  of the Royal Meteorological Society},
  \href{https://ui.adsabs.harvard.edu/abs/2005QJRMS.131..195L}{131, 195}

\bibitem[{{Lawson} {et~al.}(2019){Lawson}, {Woods}, {Jensen}, {Erfani},
  {Gurganus}, {Gallagher}, {Connolly}, {Whiteway}, {Baran}, {May},
  {Heymsfield}, {Schmitt}, {McFarquhar}, {Um}, {Protat}, {Bailey}, {Lance},
  {Muehlbauer}, {Stith}, {Korolev}, {Toon}, \& {Kr{\"a}mer}}]{lawson2019}
{Lawson}, R.~P., {Woods}, S., {Jensen}, E., {et~al.} 2019,
  \href{http://dx.doi.org/10.1029/2018JD030122}{\color{magenta}Journal of
  Geophysical Research (Atmospheres)},
  \href{https://ui.adsabs.harvard.edu/abs/2019JGRD..12410049L}{124, 10,049}

\bibitem[{{Lenoir}(1967)}]{Lenoir1967}
{Lenoir}, W.~B. 1967,
  \href{http://dx.doi.org/10.1063/1.1709315}{\color{magenta}Journal of Applied
  Physics}, \href{https://ui.adsabs.harvard.edu/abs/1967JAP....38.5283L}{38,
  5283}

\bibitem[{{Lenoir}(1968)}]{Lenoir1968}
{Lenoir}, W.~B. 1968,
  \href{http://dx.doi.org/10.1029/JA073i001p00361}{\color{magenta}\jgr},
  \href{https://ui.adsabs.harvard.edu/abs/1968JGR....73..361L}{73, 361}

\bibitem[{{Li} {et~al.}(2023){Li}, {Eimer}, {Osumi}, {Appel}, {Brewer}, {Ali},
  {Bennett}, {Bruno}, {Bustos}, {Chuss}, {Cleary}, {Couto}, {Dahal}, {Datta},
  {Denis}, {Dunner}, {Espinoza Inostroza}, {Essinger-Hileman}, {Fluxa},
  {Harrington}, {Iuliano}, {Karakla}, {Marriage}, {Miller}, {Novack},
  {N{\'u}{\~n}ez}, {Petroff}, {Reeves}, {Rostem}, {Shi}, {Valle}, {Watts},
  {Weiland}, {Wollack}, {Xu}, \& {Zeng}}]{Li23}
{Li}, Y., {Eimer}, J., {Osumi}, K., {et~al.} 2023,
  \href{https://ui.adsabs.harvard.edu/abs/2023arXiv230501045L}{\href{http://dx.doi.org/10.48550/arXiv.2305.01045}{\color{magenta}arXiv
  e-prints}, arXiv:2305.01045}

\bibitem[{Libbrecht(2017)}]{libbrecht2017physical}
Libbrecht, K.~G. 2017, Annual Review of Materials Research, 47, 271

\bibitem[{Magono(1953)}]{magono1953}
Magono, C. 1953, Sci. Rep. Yokohama Nat. Univ., 18

\bibitem[{{McFarquhar} \& {Heymsfield}(1997)}]{McFarquhar&Heymsfield1997}
{McFarquhar}, G.~M. \& {Heymsfield}, A.~J. 1997,
  \href{http://dx.doi.org/10.1175/1520-0469(1997)054<2187:POTCIC>2.0.CO;2}{\color{magenta}Journal
  of Atmospheric Sciences},
  \href{https://ui.adsabs.harvard.edu/abs/1997JAtS...54.2187M}{54, 2187}

\bibitem[{{Molod} {et~al.}(2015){Molod}, {Takacs}, {Suarez}, \&
  {Bacmeister}}]{merra2}
{Molod}, A., {Takacs}, L., {Suarez}, M., \& {Bacmeister}, J. 2015,
  \href{http://dx.doi.org/10.5194/gmd-8-1339-2015}{\color{magenta}Geoscientific
  Model Development},
  \href{https://ui.adsabs.harvard.edu/abs/2015GMD.....8.1339M}{8, 1339}

\bibitem[{{Mommert}(2020)}]{Mommert2020}
{Mommert}, M. 2020,
  \href{http://dx.doi.org/10.3847/1538-3881/ab744f}{\color{magenta}\aj},
  \href{https://ui.adsabs.harvard.edu/abs/2020AJ....159..178M}{159, 178}

\bibitem[{{Morris} {et~al.}(2022){Morris}, {Bustos}, {Calabrese}, {Choi},
  {Duivenvoorden}, {Dunkley}, {D{\"u}nner}, {Gallardo}, {Hasselfield},
  {Hincks}, {Mroczkowski}, {Naess}, {Niemack}, {Page}, {Partridge}, {Salatino},
  {Staggs}, {Treu}, {Wollack}, \& {Xu}}]{Morris2021}
{Morris}, T.~W., {Bustos}, R., {Calabrese}, E., {et~al.} 2022,
  \href{http://dx.doi.org/10.1103/PhysRevD.105.042004}{\color{magenta}\prd},
  \href{https://ui.adsabs.harvard.edu/abs/2022PhRvD.105d2004M}{105, 042004}

\bibitem[{{Nagy} {et~al.}(2017){Nagy}, {Ade}, {Amiri}, {Benton}, {Bergman},
  {Bihary}, {Bock}, {Bond}, {Bryan}, {Chiang}, {Contaldi}, {Dor{\'e}},
  {Duivenvoorden}, {Eriksen}, {Farhang}, {Filippini}, {Fissel}, {Fraisse},
  {Freese}, {Galloway}, {Gambrel}, {Gandilo}, {Ganga}, {Gudmundsson},
  {Halpern}, {Hartley}, {Hasselfield}, {Hilton}, {Holmes}, {Hristov}, {Huang},
  {Irwin}, {Jones}, {Kuo}, {Kermish}, {Li}, {Mason}, {Megerian}, {Moncelsi},
  {Morford}, {Netterfield}, {Nolta}, {Padilla}, {Racine}, {Rahlin},
  {Reintsema}, {Ruhl}, {Runyan}, {Ruud}, {Shariff}, {Soler}, {Song},
  {Trangsrud}, {Tucker}, {Tucker}, {Turner}, {Van Der List}, {Weber}, {Wehus},
  {Wiebe}, \& {Young}}]{nagy17}
{Nagy}, J.~M., {Ade}, P.~A.~R., {Amiri}, M., {et~al.} 2017,
  \href{http://dx.doi.org/10.3847/1538-4357/aa7cfd}{\color{magenta}\apj},
  \href{https://ui.adsabs.harvard.edu/abs/2017ApJ...844..151N}{844, 151}

\bibitem[{{Noel} \& {Chepfer}(2010)}]{Noel&Chepfer2010}
{Noel}, V. \& {Chepfer}, H. 2010,
  \href{http://dx.doi.org/10.1029/2009JD012365}{\color{magenta}Journal of
  Geophysical Research (Atmospheres)},
  \href{https://ui.adsabs.harvard.edu/abs/2010JGRD..115.0H23N}{115, D00H23}

\bibitem[{{Ono}(1969)}]{ono1969}
{Ono}, A. 1969,
  \href{http://dx.doi.org/10.1175/1520-0469(1969)026<0138:TSARPO>2.0.CO;2}{\color{magenta}Journal
  of Atmospheric Sciences},
  \href{https://ui.adsabs.harvard.edu/abs/1969JAtS...26..138O}{26, 138}

\bibitem[{{Padilla} {et~al.}(2020){Padilla}, {Eimer}, {Li}, {Addison}, {Ali},
  {Appel}, {Bennett}, {Bustos}, {Brewer}, {Chan}, {Chuss}, {Cleary}, {Couto},
  {Dahal}, {Denis}, {D{\"u}nner}, {Essinger-Hileman}, {Flux{\'a}}, {Gothe},
  {Haridas}, {Harrington}, {Iuliano}, {Karakla}, {Marriage}, {Miller},
  {N{\'u}{\~n}ez}, {Parker}, {Petroff}, {Reeves}, {Rostem}, {Stevens}, {Nunes
  Valle}, {Watts}, {Weiland}, {Wollack}, \& {Xu}}]{padilla20}
{Padilla}, I.~L., {Eimer}, J.~R., {Li}, Y., {et~al.} 2020,
  \href{http://dx.doi.org/10.3847/1538-4357/ab61f8}{\color{magenta}\apj},
  \href{https://ui.adsabs.harvard.edu/abs/2020ApJ...889..105P}{889, 105}

\bibitem[{Paine(2022)}]{am}
Paine, S. 2022, The am atmospheric model

\bibitem[{Paszke {et~al.}(2019)Paszke, Gross, Massa, Lerer, Bradbury, Chanan,
  Killeen, Lin, Gimelshein, Antiga, Desmaison, Kopf, Yang, DeVito, Raison,
  Tejani, Chilamkurthy, Steiner, Fang, Bai, \& Chintala}]{pytorch}
Paszke, A., Gross, S., Massa, F., {et~al.} 2019, 8024

\bibitem[{{Petroff} {et~al.}(2020){Petroff}, {Eimer}, {Harrington}, {Ali},
  {Appel}, {Bennett}, {Brewer}, {Bustos}, {Chan}, {Chuss}, {Cleary}, {Couto},
  {Dahal}, {D{\"u}nner}, {Essinger-Hileman}, {Rojas}, {Gothe}, {Iuliano},
  {Marriage}, {Miller}, {N{\'u}{\~n}ez}, {Padilla}, {Parker}, {Reeves},
  {Rostem}, {Nunes Valle}, {Watts}, {Weiland}, {Wollack}, \& {Xu}}]{petroff20}
{Petroff}, M.~A., {Eimer}, J.~R., {Harrington}, K., {et~al.} 2020,
  \href{http://dx.doi.org/10.3847/1538-4357/ab64e2}{\color{magenta}\apj},
  \href{https://ui.adsabs.harvard.edu/abs/2020ApJ...889..120P}{889, 120}

\bibitem[{{Pietranera} {et~al.}(2007){Pietranera}, {Buehler}, {Calisse},
  {Emde}, {Hayton}, {Oommen John}, {Maffei}, {Piccirillo}, {Pisano}, {Savini},
  \& {Sreerekha}}]{Pietranera2007}
{Pietranera}, L., {Buehler}, S.~A., {Calisse}, P.~G., {et~al.} 2007,
  \href{http://dx.doi.org/10.1111/j.1365-2966.2007.11464.x}{\color{magenta}\mnras},
  \href{https://ui.adsabs.harvard.edu/abs/2007MNRAS.376..645P}{376, 645}

\bibitem[{Prahl(2023)}]{miepython}
Prahl, S. 2023, {miepython: Pure python calculation of Mie scattering}

\bibitem[{{Prigent} {et~al.}(2005){Prigent}, {Defer}, {Pardo}, {Pearl},
  {Rossow}, \& {Pinty}}]{Prigent+2005}
{Prigent}, C., {Defer}, E., {Pardo}, J.~R., {et~al.} 2005,
  \href{http://dx.doi.org/10.1029/2004GL022225}{\color{magenta}\grl},
  \href{https://ui.adsabs.harvard.edu/abs/2005GeoRL..32.4810P}{32, L04810}

\bibitem[{Slonaker {et~al.}(2005)Slonaker, Takano, Liou, \& Ou}]{slonaker05}
Slonaker, R.~L., Takano, Y., Liou, K.-N., \& Ou, S.-C. 2005, 5890, 74

\bibitem[{{Stillwell} {et~al.}(2019){Stillwell}, {Neely}, {Thayer}, {Walden},
  {Shupe}, \& {Miller}}]{Stillwell2019}
{Stillwell}, R.~A., {Neely}, R.~R., {Thayer}, J.~P., {et~al.} 2019,
  \href{http://dx.doi.org/10.1029/2018JD028963}{\color{magenta}Journal of
  Geophysical Research (Atmospheres)},
  \href{https://ui.adsabs.harvard.edu/abs/2019JGRD..12412141S}{124, 12,141}

\bibitem[{{Takakura} {et~al.}(2019){Takakura}, {Aguilar-Fa{\'u}ndez}, {Akiba},
  {Arnold}, {Baccigalupi}, {Barron}, {Beck}, {Bianchini}, {Boettger},
  {Borrill}, {Cheung}, {Chinone}, {Elleflot}, {Errard}, {Fabbian}, {Feng},
  {Goeckner-Wald}, {Hamada}, {Hasegawa}, {Hazumi}, {Howe}, {Kaneko},
  {Katayama}, {Keating}, {Keskitalo}, {Kisner}, {Krachmalnicoff}, {Kusaka},
  {Lee}, {Lowry}, {Matsuda}, {May}, {Minami}, {Navaroli}, {Nishino},
  {Piccirillo}, {Poletti}, {Puglisi}, {Reichardt}, {Segawa}, {Silva-Feaver},
  {Siritanasak}, {Suzuki}, {Tajima}, {Takatori}, {Tanabe}, {Teply}, \&
  {Tsai}}]{Takakura2019}
{Takakura}, S., {Aguilar-Fa{\'u}ndez}, M.~A.~O., {Akiba}, Y., {et~al.} 2019,
  \href{http://dx.doi.org/10.3847/1538-4357/aaf381}{\color{magenta}\apj},
  \href{https://ui.adsabs.harvard.edu/abs/2019ApJ...870..102T}{870, 102}

\bibitem[{{Troitsky} \& {Osharin}(2000)}]{Troitsky&Osharin2000}
{Troitsky}, A.~V. \& {Osharin}, A.~M. 2000,
  \href{http://dx.doi.org/10.1007/BF02677151}{\color{magenta}Radiophysics and
  Quantum Electronics},
  \href{https://ui.adsabs.harvard.edu/abs/2000R&QE...43..356T}{43, 356}

\bibitem[{{Troitsky} {et~al.}(2003){Troitsky}, {Osharin}, {Korolev}, \&
  {Strapp}}]{Troitsky+2003}
{Troitsky}, A.~V., {Osharin}, A.~M., {Korolev}, A.~V., \& {Strapp}, J.~W. 2003,
  \href{http://dx.doi.org/https://doi.org/10.1175/1520-0469(2003)60<1608:POTMAR>2.0.CO;2}{\color{magenta}Journal
  of Atmospheric Sciences},
  \href{https://ui.adsabs.harvard.edu/abs/2003JAtS...60.1608T}{60, 1608}

\bibitem[{{Troitsky} {et~al.}(2005){Troitsky}, {Vostokov}, \&
  {Osharin}}]{Troitsky+2005}
{Troitsky}, A.~V., {Vostokov}, A.~V., \& {Osharin}, A.~M. 2005,
  \href{http://dx.doi.org/10.1007/s11141-005-0068-8}{\color{magenta}Radiophysics
  and Quantum Electronics},
  \href{https://ui.adsabs.harvard.edu/abs/2005R&QE...48..281T}{48, 281}

\bibitem[{Turner {et~al.}(2016)Turner, Kneifel, \&
  Cadeddu}]{turner2016improved}
Turner, D., Kneifel, S., \& Cadeddu, M. 2016, Journal of Atmospheric and
  Oceanic Technology, 33, 33

\bibitem[{{Virtanen} {et~al.}(2020){Virtanen}, {Gommers}, {Oliphant},
  {Haberland}, {Reddy}, {Cournapeau}, {Burovski}, {Peterson}, {Weckesser},
  {Bright}, {van der Walt}, {Brett}, {Wilson}, {Millman}, {Mayorov}, {Nelson},
  {Jones}, {Kern}, {Larson}, {Carey}, {Polat}, {Feng}, {Moore}, {VanderPlas},
  {Laxalde}, {Perktold}, {Cimrman}, {Henriksen}, {Quintero}, {Harris},
  {Archibald}, {Ribeiro}, {Pedregosa}, {van Mulbregt}, \& {SciPy 1. 0
  Contributors}}]{scipy}
{Virtanen}, P., {Gommers}, R., {Oliphant}, T.~E., {et~al.} 2020,
  \href{http://dx.doi.org/10.1038/s41592-019-0686-2}{\color{magenta}Nature
  Methods}, \href{https://ui.adsabs.harvard.edu/abs/2020NatMe..17..261V}{17,
  261}

\bibitem[{{Vonnegut}(1965)}]{Vonnegut1965}
{Vonnegut}, B. 1965,
  \href{http://dx.doi.org/10.1002/j.1477-8696.1965.tb02740.x}{\color{magenta}Weather},
  \href{https://ui.adsabs.harvard.edu/abs/1965Wthr...20..310V}{20, 310}

\bibitem[{{Warren} \& {Brandt}(2008)}]{Warren&Brandt2008}
{Warren}, S.~G. \& {Brandt}, R.~E. 2008,
  \href{http://dx.doi.org/10.1029/2007JD009744}{\color{magenta}Journal of
  Geophysical Research (Atmospheres)},
  \href{https://ui.adsabs.harvard.edu/abs/2008JGRD..11314220W}{113, D14220}

\bibitem[{{Westbrook} {et~al.}(2010){Westbrook}, {Illingworth}, {O'Connor}, \&
  {Hogan}}]{westbrook2010doppler}
{Westbrook}, C.~D., {Illingworth}, A.~J., {O'Connor}, E.~J., \& {Hogan}, R.~J.
  2010, \href{http://dx.doi.org/10.1002/qj.528}{\color{magenta}Quarterly
  Journal of the Royal Meteorological Society},
  \href{https://ui.adsabs.harvard.edu/abs/2010QJRMS.136..260W}{136, 260}

\bibitem[{{Zeng} {et~al.}(2023){Zeng}, {Ulanowski}, {Heymsfield}, {Wang}, \&
  {Li}}]{zeng2023}
{Zeng}, X., {Ulanowski}, Z., {Heymsfield}, A.~J., {Wang}, Y., \& {Li}, X. 2023,
  \href{http://dx.doi.org/10.1175/JAS-D-22-0223.1}{\color{magenta}Journal of
  the Atmospheric Sciences},
  \href{https://ui.adsabs.harvard.edu/abs/2023JAtS...80.1621Z}{80, 1621}

\bibitem[{{Zhang} {et~al.}(2018){Zhang}, {Liu}, {Zhang}, \&
  {Song}}]{Zhang+2018}
{Zhang}, J., {Liu}, P., {Zhang}, F., \& {Song}, Q. 2018,
  \href{http://dx.doi.org/10.1029/2018GL077787}{\color{magenta}\grl},
  \href{https://ui.adsabs.harvard.edu/abs/2018GeoRL..45.8665Z}{45, 8665}

\end{thebibliography}
\end{document}